\documentclass[a4paper]{JHEP3} 
\usepackage{epsfig,multicol,bbm}

\usepackage{ifthen}
\usepackage{epsfig}
\usepackage{amssymb}
\usepackage{graphicx}
\usepackage{amsmath}
\usepackage[center]{subfigure}
\renewcommand{\text}[1]{%
\ifthenelse{\equal{#1}{r1}}{r_1}{}%
\ifthenelse{\equal{#1}{r2}}{r_2}{}%
\ifthenelse{\equal{#1}{fB}}{f_B}{}%
\ifthenelse{\equal{#1}{fP}}{f_P}{}%
\ifthenelse{\equal{#1}{mB}}{m_B}{}%
\ifthenelse{\equal{#1}{xi0}}{\sigma_0}{}%
\ifthenelse{\equal{#1}{s0}}{s_0}{}%
\ifthenelse{\equal{#1}{M2}}{M^2}{}%
\ifthenelse{\equal{#1}{barxi}}{\bar\sigma}{}%
\ifthenelse{\equal{#1}{uu}}{u}{}%
\ifthenelse{\equal{#1}{ubar}}{\bar{\sigma}}{}%
\ifthenelse{\equal{#1}{vv}}{v}{}
\ifthenelse{\equal{#1}{phiBp}}{\phi^B_+}{}%
\ifthenelse{\equal{#1}{PhiBp}}{\phi^B_+}{}%
\ifthenelse{\equal{#1}{phiBmin}}{\phi^B_-}{}%
\ifthenelse{\equal{#1}{PhiBmin}}{\phi^B_-}{}%
\ifthenelse{\equal{#1}{PhiBpm}}{\overline\Phi^B_{\pm}}{}%
\ifthenelse{\equal{#1}{ppsiV}}{\Psi_V^B}{}%
\ifthenelse{\equal{#1}{ppsiA}}{(\Psi_A^B}{}%
\ifthenelse{\equal{#1}{bbarXA}}{\Psi_V^B}{}%
\ifthenelse{\equal{#1}{bbarYA}}{\overline Y_A^B}{}%
\ifthenelse{\equal{#1}{fV}}{f_V}{}%
\ifthenelse{\equal{#1}{mV}}{m_V}{}%
\ifthenelse{\equal{#1}{mP}}{m_P}{}%
\ifthenelse{\equal{#1}{lb}}{\lambda_B}{}%
}

\newcommand{\ba}{\begin{eqnarray}}
\newcommand{\ea}{\end{eqnarray}}
\newcommand{\be}{\begin{equation}}
\newcommand{\ee}{\end{equation}}
\renewcommand{\log}{\ln}
\newcommand{\DS}[1]{/\!\!\!#1}

\makeatletter
\def\fmslash{\@ifnextchar[{\fmsl@sh}{\fmsl@sh[0mu]}}
\def\fmsl@sh[#1]#2{%
  \mathchoice
    {\@fmsl@sh\displaystyle{#1}{#2}}%
    {\@fmsl@sh\textstyle{#1}{#2}}%
    {\@fmsl@sh\scriptstyle{#1}{#2}}%
    {\@fmsl@sh\scriptscriptstyle{#1}{#2}}}
\def\@fmsl@sh#1#2#3{\m@th\ooalign{$\hfil#1\mkern#2/\hfil$\crcr$#1#3$}}
\makeatother

\title{Light-cone sum rules for
$B\to \pi$ form factors  revisited}

\author{G.~Duplan\v ci\'c\thanks{Alexander von Humboldt Fellow}
\\
Max-Planck-Institut fur Physik
(Werner-Heisenberg-Institut), 
F{\"o}hringer Ring 6, 
D-80805 M{\"u}nchen, Germany\\
Rudjer Boskovic Institute, Theoretical Physics Division, HR-10002 Zagreb, Croatia\\
        E-mail: \email{gorand@thphys.irb.hr}}
\author{A.~Khodjamirian\\
Theoretische Physik 1, Fachbereich Physik,
Universit\"at Siegen, D-57068 Siegen, Germany\\
        E-mail: \email{khodjam@hep.physik.uni-siegen.de}}
\author{Th.~Mannel\\
Theoretische Physik 1, Fachbereich Physik,
Universit\"at Siegen, D-57068 Siegen, Germany\\
        E-mail: \email{mannel@hep.physik.uni-siegen.de}}
\author{B.~Meli\'c\\
Rudjer Boskovic Institute, Theoretical Physics Division, HR-10002 Zagreb, Croatia\\
        E-mail: \email{melic@thphys.irb.hr}}
\author{N.~Offen\\
Laboratoire de Physique Th\'eorique
CNRS/Univ. Paris-Sud 11, F-91405 Orsay, France\\
        E-mail: \email{nils.offen@th.u-psud.fr}}

\abstract{
We reconsider and update the QCD light-cone sum rules 
for $B\to \pi$ form factors. The gluon 
radiative corrections to the twist-2 and twist-3 terms 
in the correlation functions are calculated. 
The $\overline{MS}$ $b$-quark mass is employed,  
instead of the one-loop pole mass used 
in the previous analyses.
The light-cone sum rule for $f^+_{B\pi}(q^2)$ is fitted 
to the measured $q^2$-distribution in $B\to \pi l \nu_l$, 
fixing the input parameters with the largest uncertainty:
the Gegenbauer moments of the pion distribution amplitude.  
For the $B\to \pi$ vector form factor at
zero momentum transfer we predict $f^+_{B\pi}(0)= 0.26^{+0.04}_{-0.03}$. 
Combining it with the value of the product $|V_{ub}f^+_{B\pi}(0)|$ extracted 
from experiment, we obtain $|V_{ub}|=(3.5\pm 0.4\pm 0.2\pm 0.1) \times 10^{-3}$.
In addition, the scalar and penguin $B\to \pi$ form factors
$f^0_{B\pi}(q^2)$ and $f^T_{B\pi}(q^2)$ are calculated.} 


\begin{document}

\section{Introduction} 

The form factors of heavy-to-light transitions 
at large energies of the final hadrons are 
among the most important applications 
of QCD light-cone sum rules (LCSR)  \cite{lcsr}.
In this paper we concentrate on the $B\to \pi$ transition 
form factors $f^+_{B\pi}$, 
$f^0_{B\pi}$ and $f^T_{B\pi}$ of the electroweak vector $b\to u$ and penguin $b\to d$ currents, respectively. 
Previously, these form factors have 
been calculated from LCSR in  
\cite{BKR,BBKR,KRWY,BBB,Ball98,KRW,KRWWY,BZ01,BZ04}, 
gradually improving the accuracy.

The main advantage  of 
LCSR is the possibility  to perform calculations in full QCD, with 
a finite $b$-quark mass. In the sum rule approach, the 
$B\to\pi$ matrix element is obtained from 
the correlation function
of quark currents, rather than estimated directly from a  
certain factorization ansatz. This correlation function  
is conveniently ``designed'', so that, at  
large spacelike external momenta, the operator-product expansion
(OPE) near the light-cone is applicable. Within OPE,   
the correlation function is factorized
in a series of hard-scattering amplitudes convoluted with 
the pion light-cone distribution amplitudes (DA's) of growing 
twist. To obtain the $B\to\pi$  form factors from the 
correlation function, one makes 
use of the hadronic dispersion relation and quark-hadron duality
in the $B$-meson channel,
following the general strategy of QCD sum rules \cite{SVZ}.
More details can be found 
in the reviews on LCSR, e.g.,  in \cite{KR,Braun97,CK}.  
A modification of the method, involving $B$-meson 
distribution amplitudes and dispersion relation in the pion 
channel was recently suggested in \cite{KMO};
the analogous sum rules for $B\to\pi$ form factors 
in soft-collinear effective theory (SCET) were derived in \cite{DFFH}.

LCSR provide analytic expressions for the 
form  factors, including both hard-scattering and 
soft (end-point) contributions.
Because the method is based on a calculation 
in full QCD, combined with a rigorous hadronic dispersion relation,
the uncertainties in the resulting
LCSR are identifiable and assessable. 
These uncertainties are caused by 
the truncation of the light-cone 
OPE, and by the limited accuracy of the universal input, 
such as the quark masses  
and  parameters of the pion DA's.
In addition, a sort of systematic uncertainty 
is brought by the quark-hadron duality  approximation
adopted for the contribution of excited hadronic states 
in the dispersion relation. 
Importantly, $B\to \pi$ form factors are 
calculable from LCSR in the 
region of small momentum transfer $q^2$ (large energy of the pion), 
not yet directly accessible to lattice QCD.

The $B\to \pi l\nu_l$ decays, 
with continuously improving experimental data, provide nowadays 
the most reliable exclusive $V_{ub}$ determination. 
Along with  the lattice QCD results, the form factor 
$f^+_{B\pi}(q^2)$ obtained \cite{BZ04} from LCSR is used for the $|V_{ub}|$ 
extraction. Furthermore, the LCSR 
form factors $f^{+,0}_{B\pi}(q^2)$ can provide inputs for 
various factorization 
approaches to  exclusive $B$ decays, such as 
QCD factorization \cite{BBNS}, 
whereas the penguin form factor  $f^T_{B\pi}$
is necessary for the analysis of the rare 
$B\to \pi l^+l^-$ decay. Having in mind  the importance of  
$B\to \pi$ form factors for the 
$V_{ub}$ determination and for the phenomenological analysis of 
various exclusive $B$ decays, we decided
to reanalyze and update the LCSR for these form factors. 
One of our motivations 
was to recalculate the $O(\alpha_s)$  gluon 
radiative correction to the twist-3 part of the correlation function.  
Only a single calculation 
of this term  exists \cite{BZ01,BZ04}, whereas the
$O(\alpha_s)$  corrections to the twist-2 part 
have been independently obtained in \cite{KRWY} and \cite{BBB}. 
In what follows, we derive and present the explicit expressions 
for all  $O(\alpha_s)$ hard-scattering 
amplitudes and their imaginary parts for the twist-2 and twist-3 
parts of the correlation function and some of these expressions 
are new.

In the OPE of the correlation function 
the $\overline{MS}$ mass 
$\overline{m}_b(\mu)$ is used, a natural choice 
for a virtual $b$-quark propagating in the 
hard-scattering amplitudes, calculated at large spacelike 
momentum scales $\sim m_b$. 
Importantly, in the resulting sum rules we keep using
the $\overline{MS}$ mass. Note that the value of 
$\overline{m}_b(\overline{m}_b)$  
is rather accurately determined from the bottomonium sum rules.  
In previous analyses, the one-loop pole mass of the $b$-quark 
was employed in LCSR. The main motivation was 
that the pole mass was used also in the two-point sum rule for 
the $B$-meson decay constant
$f_B$, needed to extract the form factor from LCSR. 
In the meantime,  the $f_B$ sum rule  is available 
also in $\overline{MS}$-scheme  \cite{JL}, and  we apply  
this new version here.

Furthermore, we fix the most uncertain 
input parameters, the effective threshold and 
simultaneously, the Gegenbauer moments 
of the pion twist-2 DA, 
by calculating the $B$-meson mass and the shape of $f^+_{B\pi}(q^2)$
from LCSR and 
fitting these quantities to their measured values.
In addition, the nonperturbative 
parameters of the twist-3,4 pion DA's entering LCSR are updated,
using the results of  the recent analysis \cite{BBL}. 
 
The paper is organized as follows. 
In sect.~2 the correlation function is introduced and 
the leading-order (LO) terms of OPE are presented, 
including  the contributions 
of the pion twist-2,3,4 two-particle DA's and twist-3,4 three-particle DA's. 
In sect.~3 the calculation of the $O(\alpha_s)$ 
twist-2 and  twist-3 parts of the correlation 
function is discussed. 
In sect. 4  we present LCSR for all three $B\to \pi$ 
form factors. Sect. 5 contains the discussion of the numerical input 
and results, as well as the estimation of theoretical uncertainties,
and finally, the determination of $|V_{ub}|$. 
Sect. 6 is devoted to the concluding discussion. 
App.~A contains the 
necessary formulae and input for the pion DA's. 
The bulky expressions for the $O(\alpha_s)$ hard-scattering amplitudes 
and their imaginary parts  are collected in App.~B,
and the sum rule for $f_B$ is given in App.~C.

\section{Correlation function}

The vacuum-to-pion correlation function used to obtain the LCSR 
for the form factors of $B\to\pi$ transitions is defined as:
\ba
F_{\mu}(p,q)&=&i\int d^4x ~e^{i q\cdot x}
\langle \pi^+(p)|T\left\{\bar{u}(x)\Gamma_\mu b(x), 
m_b\bar{b}(0)i\gamma_5 d(0)
\right\}|0\rangle
\nonumber\\
&=&\Bigg\{\begin{array}{ll}
F(q^2,(p+q)^2)p_\mu +\widetilde{F}(q^2,(p+q)^2)q_\mu\,,& ~~\Gamma_\mu= \gamma_\mu\\
&\\
F^T(q^2,(p+q)^2)\big[p_\mu q^2-q_\mu (q p)\big]\,,& ~~\Gamma_\mu= -i\sigma_{\mu\nu}q^\nu\\
\end{array} 
\label{eq:corr}
\ea
for  the two different $b\to u$ transition currents, 
For definiteness, we consider the 
$\bar{B}_d\to \pi^+$ flavour configuration
and, for simplicity we use $u$ instead of $d$ in the penguin
current, which does not make difference  
in the adopted isospin symmetry limit.
Working in  the chiral limit, we neglect the pion mass 
($p^2=m_\pi^2=0$)  
and the $u$-, $d$-quark masses, whereas the ratio 
$\mu_\pi=m_\pi^2/(m_u+m_d)$ remains  finite.

At $q^2\ll m_b^2$ and $(p+q)^2\ll m_b^2$, that is, 
far from the $b$-flavour thresholds, 
the $b$ quark propagating in the correlation 
function is highly virtual and the distances near the light-cone $x^2= 0$ 
dominate.
It is possible to prove the light-cone dominance,
following the same line of arguments as in \cite{KMO}. 
Contracting the $b$-quark fields, one expands the vacuum-to pion 
matrix element in terms of the pion light-cone DA's of growing twist.
The light-cone expansion \cite{BB} of the $b$-quark propagator 
is used (see also \cite{BBKR}):
\ba
&& \langle 0 |b^i_\alpha(x)\bar{b}^j_\beta(0)|0 \rangle = 
-i\int\frac{d^4k}{(2\pi)^4}
e^{-ik\cdot x}\Bigg[ \delta^{ij}\frac{\DS k+m }{m^2-k^2}
\nonumber\\
&& \hspace*{1cm}+g_s\int\limits_0^1 dv G^{\mu\nu a}(vx)
\left(\frac{\lambda^{a}}2\right)^{ij}
\Bigg(\frac{\DS k+m}{2(m^2-k^2)^2}\,\sigma_{\mu\nu}+
\frac{1}{m^2-k^2}\, vx_\mu\gamma_\nu \Bigg)\Bigg]_{\alpha\beta}\,,
\label{eq:prop}
\ea
where only the free propagator and the one-gluon 
term are retained. The latter  term gives rise to
 the three-particle DA's in the OPE.
\FIGURE[t]{
\includegraphics[width=5cm]{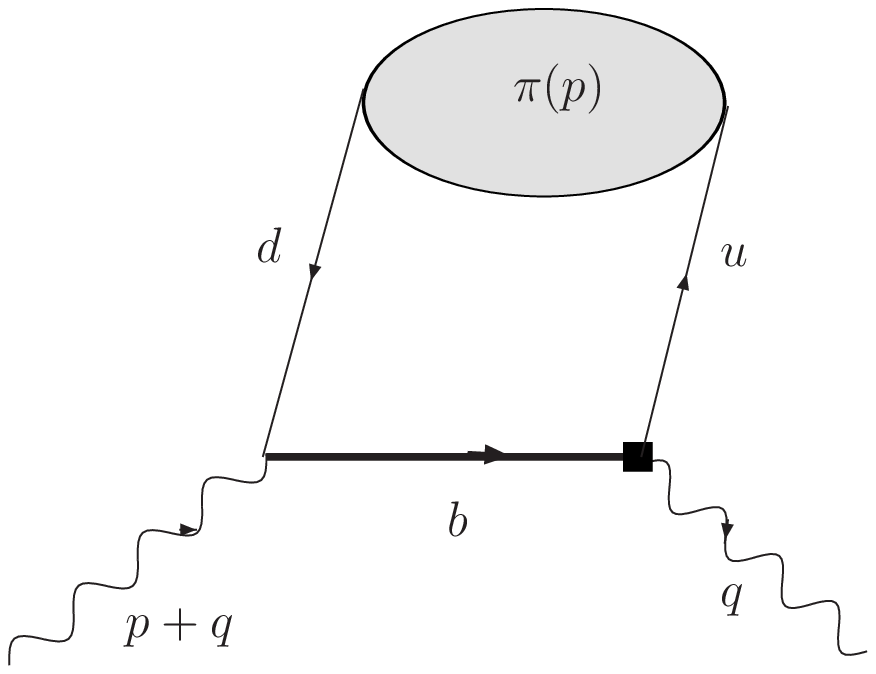}\hspace{1cm}
\includegraphics[width=5cm]{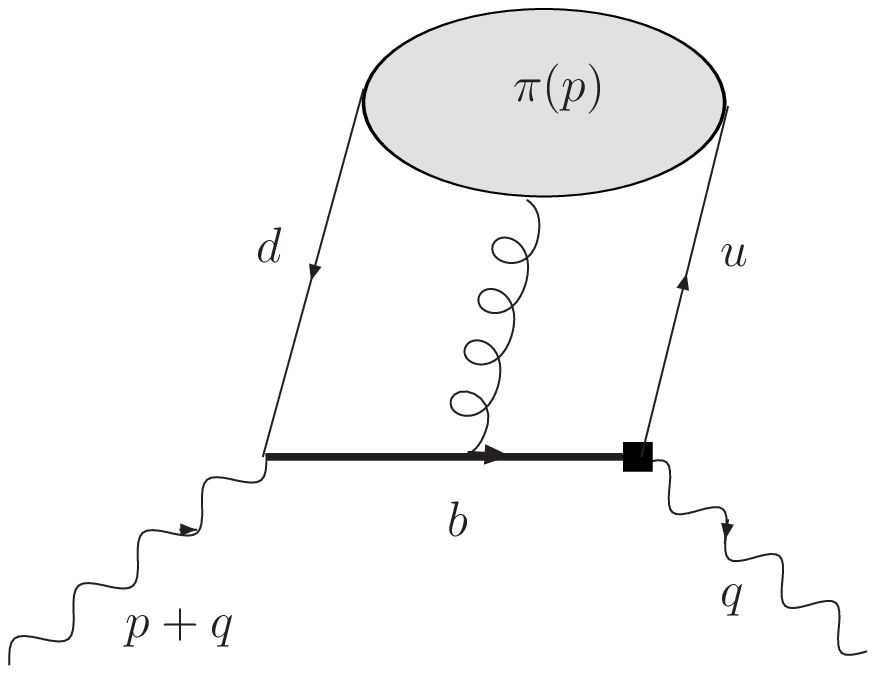}\\
\caption{ \it Diagrams representing the leading-order terms in 
the correlation function involving the  
two-particle (left) and three-particle (right) pion DA's shown by ovals. 
Solid, curly and wave lines represent quarks, gluons, and external 
currents, respectively.}
\label{fig-diags}
}
Diagrammatically, the contributions of two- and three-particle
DA's to the correlation function are depicted in Fig.~\ref{fig-diags}.
In terms of perturbative QCD, these
are LO (zeroth order in $\alpha_s$) contributions.  
The Fock components of the pion with multiplicities 
larger than three, 
are neglected, as well as the twists higher than 4. 
This truncation
is justified by the fact that the twist-4 and three-particle
corrections to LCSR obtained below turn out to be very small.   

In addition we include the $O(\alpha_s)$ 
gluon radiative corrections 
to the dominant twist-2 and twist-3 parts  of
the correlation function. The OPE result for 
the invariant amplitude $F$  is then represented as
a sum of LO and NLO parts:   
\be 
F(q^2,(p+q)^2)= F_0(q^2,(p+q)^2)+\frac{\alpha_sC_F}{4\pi}F_1(q^2,(p+q)^2),
\label{eq:alphasexp}
\ee
and the same for $\widetilde{F}$ and $F^T$.
The leading-order (LO) invariant amplitudes $F_0$, 
$\widetilde{F}_0$, and $F^T_0$ 
including twist 2,3,4 contributions have been obtained earlier
in \cite{BBKR,Ball98,KRW,Aliev}.
We present them  here switching to the new notations \cite{BBL} 
of the twist-3,4 DA's:
\ba
F_0(q^2,(p+q)^2)&=& m_b^2 f_\pi \int \limits_0^1\frac{du}{m_b^2-(q+up)^2}\Bigg\{
\varphi_\pi(u)
+\frac{\mu_\pi}{m_b}u\phi^p_{3\pi}(u)
\nonumber
\\
& & +\frac{\mu_\pi}{6m_b}
\Bigg[ 2+\frac{m_b^2+q^2}{m_b^2-(q+up)^2}\Bigg]\phi^\sigma_{3\pi}(u)
-\frac{m_b^2\phi_{4\pi}(u)}{2\big(m_b^2-(q+up)^2\big)^2}
\nonumber \\
&& -\frac{u}{m_b^2-(q+up)^2}\int\limits_0^u dv \psi_{4\pi}(v)\Bigg\}
\nonumber
\\
& & +\int\limits _0^1 dv \int 
\frac{{\cal D}\alpha}{\big[m_b^2-
\big(q+(\alpha_1+\alpha_3v)p\big)^2\big]^2}\Bigg\{
4 m_b f_{3\pi} v(q\cdot p)\Phi_{3\pi}(\alpha_i)
\nonumber
\\
& & +m_b^2 f_\pi \bigg( 2\Psi_{4\pi}(\alpha_i)-\Phi_{4\pi}(\alpha_i)+
2\widetilde{\Psi}_{4\pi}(\alpha_i)-
\widetilde{\Phi}_{4\pi}(\alpha_i)\bigg)\Bigg\}\,,
\label{eq:corrF}
\ea

\ba
\widetilde{F}_0(q^2,(p+q)^2)&=& m_b f_\pi\int \limits_0^1\frac{du}{m_b^2-(q+up)^2}\Bigg\{
\mu_\pi\phi^p_{3\pi}(u) \qquad\qquad
\nonumber
\\
&& +\frac{\mu_\pi}{6}
\Bigg[ 1-\frac{m_b^2-q^2}{m_b^2-(q+up)^2}\Bigg]
\frac{\phi^\sigma_{3\pi}(u)}{u}
-\frac{m_b}{m_b^2-(q+up)^2}\int\limits_0^u dv \psi_{4\pi}(v)\Bigg\}\,,
\nonumber \\
&&
\label{eq:corrFtilde}
\ea
\ba
F^T_0(q^2,(p+q)^2)= m_b f_\pi\int \limits_0^1\frac{du}{m_b^2-(q+up)^2}\Bigg\{
\varphi_\pi(u)
+\frac{m_b\mu_\pi}{3(m_b^2-(q+up)^2)}\phi^\sigma_{3\pi}(u)
\nonumber
\\
-\frac1{2(m_b^2-(q+up)^2)}\Bigg(\frac12 +
\frac{m_b^2}{m_b^2-(q+up)^2}\Bigg)\phi_{4\pi}(u)\Bigg\}
\nonumber
\\
+m_b f_\pi \int\limits _0^1 dv \int 
\frac{{\cal D}\alpha}{\big[m_b^2-
\big(q+(\alpha_1+\alpha_3v)p\big)^2\big]^2}
\Bigg\{2\Psi_{4\pi}(\alpha_i)-(1-2v)\Phi_{4\pi}(\alpha_i)
\nonumber
\\
+
2(1-2v)\widetilde{\Psi}_{4\pi}(\alpha_i)-
\widetilde{\Phi}_{4\pi}(\alpha_i)\Bigg\}\,,
\label{eq:corrFtens}
\ea
where ${\cal D}\alpha=
d\alpha_1 d\alpha_2 d\alpha_3 \delta(1-\alpha_1-\alpha_2-\alpha_3)$,
and the definitions of the twist-2 ($\varphi_\pi$), twist-3
($\phi^p_{3\pi}$, $\phi^\sigma_{3\pi}$, $\Phi_{3\pi}$) 
and twist-4 ($\phi_{4\pi}$, $\psi_{4\pi}$, 
$\Phi_{4\pi}$, $\Psi_{4\pi}$, $\widetilde{\Phi}_{4\pi}$,
$\widetilde{\Psi}_{4\pi}$) pion DA's and their parameters are presented in 
App.~A.
Note that all twist-4 terms are suppressed with respect
to leading twist-2 terms, 
with an additional power of the denominator $1/(m_b^2-(q+up)^2)$
compensated by the normalization parameter 
$\delta_\pi^2\sim \Lambda_{QCD}^2$ of the twist-4 DA's. 

The calculation of the NLO amplitudes  
$F_1,\tilde{F}_1, F^T_1$ will be discussed in the next section.
\section{Gluon radiative corrections}

\FIGURE[t]{
\includegraphics[width=14cm]{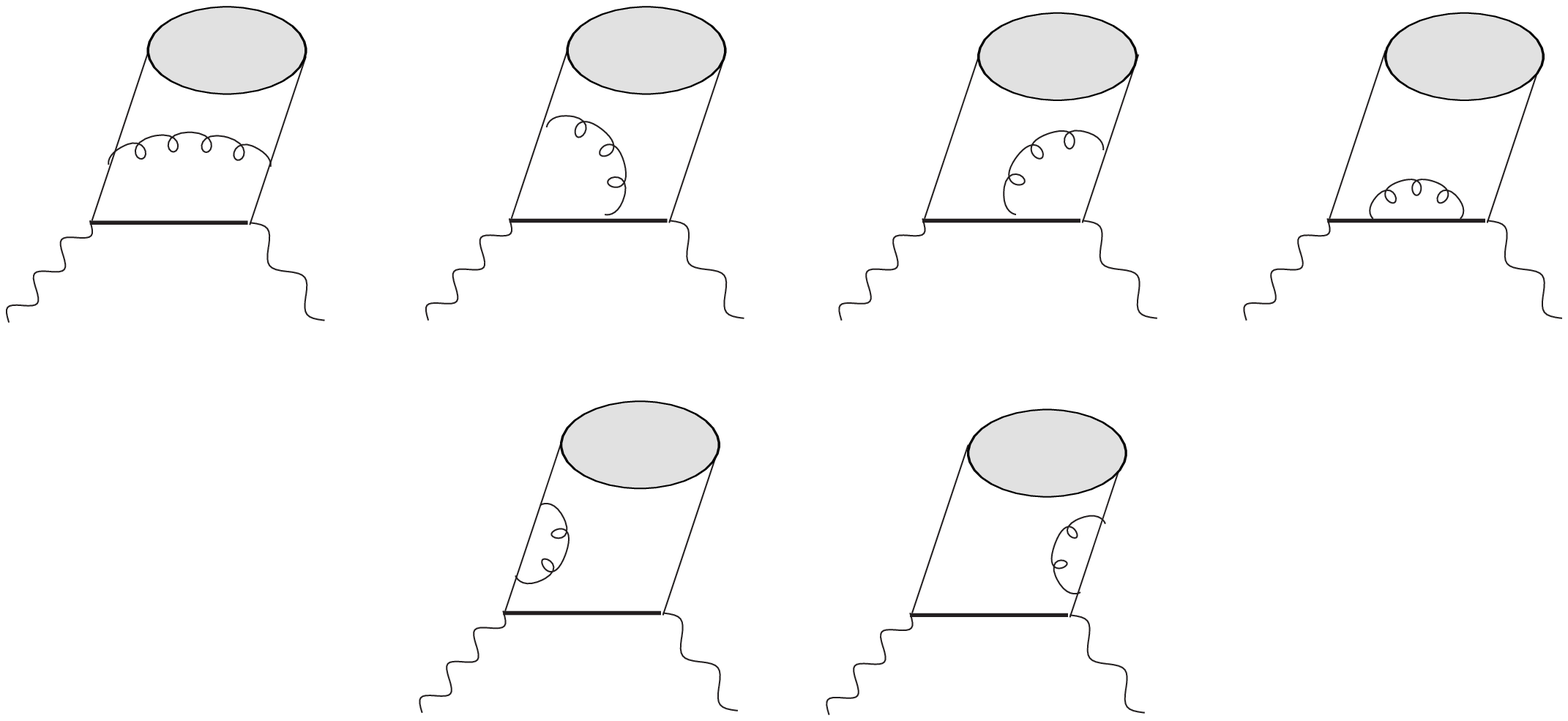}\\
\caption{ \it Diagrams corresponding to
the $O(\alpha_s)$ gluon radiative corrections to the correlation function.}
\label{fig-alphas}
}

In the light-cone OPE of 
the correlation function (\ref{eq:corr}) 
each twist component receives gluon radiative corrections. 
To obtain the desired NLO terms, one has to calculate the $O(\alpha_s)$ 
one-loop diagrams shown in Fig.~\ref{fig-alphas}, convoluting them with 
the twist-2 and two-particle twist-3 DA's, respectively.
The diagrams are computed using the standard dimensional 
regularization and $\overline{MS}$ scheme. 
In addition, in our calculation the reduction method 
from \cite{DN} is employed.

The invariant amplitude 
$F_1$ in (\ref{eq:alphasexp}) is obtained 
in a factorized form of the convolutions: 
\ba
F_1(q^2,(p+q)^2) &=& f_\pi \int_0^1 du
\Bigg\{ T_1(q^2,(p+q)^2,u)\varphi_\pi(u)
\nonumber\\
& & +
\frac{\mu_\pi}{m_b}\Big[
T_1^p(q^2,(p+q)^2,u)\phi^p_{3\pi}(u)+
T_1^\sigma(q^2,(p+q)^2,u)\phi^\sigma_{3\pi}(u)\Big]\Bigg\}\,,
\label{eq:convol}
\ea
where the hard-scattering amplitudes $T_1$, $T_1^{p,\sigma}$ result from the
calculation of the diagrams in Fig.\ref{fig-alphas}. The two other NLO amplitudes
 $\widetilde{F}_1$ and $m_bF_1^T$  have the same expressions 
with $T_1\to \widetilde{T}_1$, 
$T_1^{p,\sigma}\to \widetilde{T}_1^{p,\sigma}$, 
and $T_1\to T_1^T$, $T_1 \to T_1^{T p,\sigma}$, respectively. 
The resulting expressions 
for all hard-scattering amplitudes are presented in App.~B. 
Note that the LO expressions for the correlation functions
in (\ref{eq:corrF})-(\ref{eq:corrFtens}) 
also have a factorized, albeit a much simpler form, 
with the zeroth-order in $\alpha_s$ hard-scattering amplitudes 
stemming from the free propagator of the virtual $b$-quark. 
In particular, the twist-2 component in $F_0$ is 
a convolution of   
$T_0=m_b^2/[m_b^2-(q+u p)^2]$  with $\varphi_\pi(u)$.

Let us mention some important features of the $O(\alpha_s)$ 
terms of OPE. 
The currents $\bar{u}\gamma_\mu b$ and $m_b\bar{b}i\gamma_5 d$ 
in the correlation function are physical and not renormalizable.
Hence, the ultraviolet singularities appearing in $T_1$ and 
$\widetilde{T}_1$ are canceled by the 
renormalization of the heavy quark mass. For $T_1^T$ 
an additional renormalization of the composite 
$\overline{q}\sigma_{\mu\nu}b$ operator has to be taken into
account.
Furthermore, in the twist-2 term in (\ref{eq:convol}) the 
convolution integral is convergent due to collinear 
factorization. As explicitly shown in \cite{KRWY,BBB},
the infrared-collinear divergences of the $O(\alpha_s)$ 
diagrams are absorbed  by the well known 
one-loop evolution \cite{ERBL} of the twist-2 pion DA.  
As a result of factorization, a residual dependence on the 
factorization scale $\mu_f$ 
enters the amplitude $T_1$ and the twist-2 DA 
$\varphi_\pi$. This scale effectively separates the 
long- and short (near the light-cone) distances 
in the correlation function.   
In the twist-3 part of $F_1$,  the complete 
evolution kernel has to include the mixing of two- and  
three-particle DA's. To avoid these complications,
and following \cite{BZ01}, the twist-3 pion DA's 
in (\ref{eq:convol})
are taken in their asymptotic form: 
$\phi_p(u)=1$ and $\phi_{\sigma}(u)=6u(1-u)$,
whereas the nonasymptotic effects in these DA's 
are only included in  the LO part $F_0$.  
We checked that the infrared divergences appearing 
in the amplitudes $T_1^p$ and $T_1^\sigma$ 
cancel in the sum of the $\phi_p$ and $\phi_{\sigma}$ contributions 
with the one-loop renormalization of the 
parameter $\mu_{\pi}$ (i.e., of the quark condensate density). 
Finally, in accordance with 
\cite{BZ01,BZ04}, all renormalized 
hard-scattering amplitudes 
are well behaved at the end-points $u = 0,1$, 
regardless of the form of the DA's.

After completing the calculation of OPE terms with the LO (NLO) 
accuracy up to twist-4 (twist-3), we turn now to the 
derivation of the sum rules.

\section{LCSR for $B\to\pi$   form factors}
In the LCSR approach 
the  $B\to\pi$ matrix elements  
are related to the correlation function 
(\ref{eq:corr})  
via hadronic dispersion relation in the channel 
of the $\bar{b}\gamma_5 d$ 
current with the four-momentum squared $(p+q)^2$.
Inserting hadronic states between the currents 
in (\ref{eq:corr}) one isolates the 
ground-state $B$-meson contributions in the  
dispersion relations for 
all three invariant amplitudes:
\ba
&F(q^2,(p+q)^2)&=
\frac{2m_B^2f_Bf^+_{B\pi}(q^2)}{m_B^2-(p+q)^2}+\ldots
\nonumber
\\
&\widetilde{F}(q^2,(p+q)^2)&=
\frac{m_B^2 f_B[f^+_{B\pi}(q^2)+f^-_{B\pi}(q^2)]}{m_B^2-(p+q)^2}
+\ldots
\nonumber
\\
&F^T(q^2,(p+q)^2)&=\frac{2 m_B^2 f_B f_{B\pi}^T(q^2)}{(m_B+
m_\pi)(m_B^2-(p+q)^2)}+\ldots
\label{eq:disp}
\ea
where the ellipses indicate the contributions of heavier 
states (starting from $B^*\pi$). 
The three $B\to \pi$ form factors 
entering the residues of the $B$ pole in (\ref{eq:disp})
are defined as: 
\be
\langle\pi^+(p)|\bar{u} \gamma_\mu b |\bar B_d(p+q)\rangle=
2f^+_{B\pi}(q^2)p_\mu +\left(f^+_{B\pi}(q^2)+f^-_{B\pi}(q^2)\right)q_\mu\,,
\label{eq:fplBpi}
\ee
\be
\langle\pi^+(p)|\bar{u} \sigma_{\mu \nu}q^\nu b
|\bar B_d(p+q)\rangle=
\Big [q^2(2p_\mu+q_\mu) - (m_B^2-m_\pi^2) q_\mu\Big ]
\frac{i f_{B\pi}^T(q^2)}{m_B+m_\pi}\,,
\ee
and $f_B=\langle \bar{B}_d |m_b\bar{b}i\gamma_5 d |0 \rangle/m_B^2$ is the $B$-meson decay constant.

Substituting the OPE results for $F$, $\widetilde{F}$ and $F^T$ 
in l.h.s. of (\ref{eq:disp}), one approximates the contributions of 
the heavier states in r.h.s. with the help of quark-hadron duality,
introducing the effective threshold parameter $s_0^B$.
After the Borel transformation in the variable 
$(p+q)^2\to M^2$, the sum rules for all three 
$B\to \pi$ form factors are obtained. 
The LCSR for the vector form factor reads:   
\be
f^+_{B\pi}(q^2)= \frac{e^{m_B^2/M^2}}{2m_B^2 f_B} 
\Bigg[F_0(q^2,M^2,s_0^B)+
\frac{\alpha_s C_F}{4\pi}F_1(q^2,M^2,s_0^B)
\Bigg]\,,
\label{eq:fplusLCSR}
\ee
where $F_{0(1)}(q^2,M^2,s_0^B)$ originates
from the OPE result for the LO (NLO) invariant 
amplitude $F_{0(1)}(q^2,(p+q)^2)$. 

The LO part of the LCSR has the following expression:
\ba
&& F_0(q^2,M^2,s_0^B)= m_b^2f_\pi\int\limits_{u_0}^1 du\,
e^{-\frac{m_b^2-q^2\bar{u}}{uM^2}}
\Bigg\{\frac{\varphi_\pi(u)}{u} \qquad\qquad\qquad\qquad
\nonumber
\\
&& \qquad +\frac{\mu_\pi}{m_b}\Bigg(\phi_{3\pi}^p(u)
+\frac{1}{6}\Big[ \frac{2\phi_{3\pi}^\sigma(u)}{u}
-\left(\frac{m_b^2+q^2}{m_b^2-q^2}\right)\frac{d\phi_{3\pi}^{\sigma}(u)}{du}\Big]\Bigg)
-2\left(\frac{f_{3\pi}}{m_b f_\pi}\right)\frac{I_{3\pi}(u)}{u}
\nonumber
\\
& & \qquad +\frac{1}{m_b^2-q^2}
\Bigg(-\frac{m_b^2\,u}{4(m_b^2-q^2)}
\frac{d^2\phi_{4\pi}(u)}{du^2}
+u\psi_{4\pi}(u)+\int\limits_0^u dv \psi_{4\pi}(v)
-I_{4\pi}(u)\Bigg)\Bigg\},
\label{eq:fplusBpiLCSRcontrib}
\ea
where $\bar{u}=1-u$, $u_0=(m_b^2-q^2)/(s_0^B-q^2)$  and the 
short-hand notations
introduced for the integrals over three-particle DA's are:
\ba
&& I_{3\pi}(u)=\frac{d}{du}\Bigg(\int\limits_0^u \!d\alpha_1\!\!\!
\int\limits_{(u-\alpha_1)/(1-\alpha_1)}^1\!\!\!\!\! dv \,\,\Phi_{3\pi}(\alpha_i)
\Bigg|_{\begin{array}{l}
\alpha_2=1-\alpha_1-\alpha_3,\\
\alpha_3=(u-\alpha_1)/v
\end{array}
}
\Bigg)
\,,
\nonumber
\\
&& I_{4\pi}(u)=\frac{d}{du}\Bigg(\int\limits_0^u\! d\alpha_1\!\!\!
\int\limits_{(u-\alpha_1)/(1-\alpha_1)}^1\!\!\!\!\! \frac{dv}{v} \,\,
\Bigg[2\Psi_{4\pi}(\alpha_i)-\Phi_{4\pi}(\alpha_i)
\nonumber \\
&& \qquad\qquad\qquad\qquad +
2\widetilde{\Psi}_{4\pi}(\alpha_i)-
\widetilde{\Phi}_{4\pi}(\alpha_i)\Bigg]
\Bigg|_{\begin{array}{l}
\alpha_2=1-\alpha_1-\alpha_3,\\
\alpha_3=(u-\alpha_1)/v
\end{array}
}\Bigg)\,.
\label{eq:fplusBpiLCSR3part}
\ea
The NLO term in (\ref{eq:fplusLCSR}) 
is cast in the form of the dispersion relation:
\ba
&&F_1(q^2,M^2,s_0^B) = \frac{1}{\pi}\int\limits_{m_b^2}^{s_0^B}
ds e^{-s/M^2}\,\mbox{Im}_s F_1(q^2,s) 
\nonumber\\
&& \qquad\qquad = \frac{f_\pi}{\pi} \int\limits_{m_b^2}^{s_0^B}ds e^{-s/M^2}
\int_0^1 du\Bigg\{\mbox{Im}_sT_1(q^2,s,u)\,\varphi_\pi(u)\ 
\nonumber\\
&& \qquad\qquad\qquad +
\frac{\mu_\pi}{m_b}\Big[
\,\mbox{Im}_sT_1^p(q^2,s,u)\,\phi^p_{3\pi}(u)\,+
\,\mbox{Im}_sT_1^\sigma(q^2,s,u)\,\phi^\sigma_{3\pi}(u)\Big]\Bigg\}\,,
\label{eq:Imconvol}
\ea
where the bulky expressions for the imaginary parts of the 
amplitudes $T_1$,$T_1^p$,$T_1^\sigma$  are presented in App.~B.

The LCSR following from the dispersion relation 
for the invariant amplitude $\widetilde{F}$ in (\ref{eq:disp}) reads:
\ba
f^+_{B\pi}(q^2)+f^-_{B\pi}(q^2)= 
\frac{e^{m_B^2/M^2}}{m_B^2 f_B}
\Bigg[\widetilde{F}_0(q^2,M^2,s_0^B)+
\frac{\alpha_s C_F}{4\pi}\widetilde{F}_1(q^2,M^2,s_0^B)
\Bigg]\,,
\label{eq:fplminLCSR}
\ea
where 
\ba
\widetilde{F}_0(q^2,M^2,s_0^B)= 
m_b^2 f_\pi\int\limits_{u_0}^1 du\,
e^{-\frac{m_b^2-q^2\bar{u}}{uM^2}}
\Bigg\{\frac{\mu_\pi}{m_b}\Bigg(\frac{\phi_{3\pi}^p(u)}{u}
+\frac{1}{6u}\frac{d\phi_{3\pi}^{\sigma}(u)}{du}\Bigg)
\nonumber
\\
+\frac{1}{m_b^2-q^2}\psi_{4\pi}(u)
\Bigg\}\,.
\label{eq:fplminBpiLCSRcontrib}
\ea
Here the contributions of twist-2 and of three-particle DA's vanish
altogether.
Combining (\ref{eq:fplusLCSR}) and  (\ref{eq:fplminLCSR}) 
one is able to calculate the scalar $B\to \pi$ form factor:
\be
f^0_{B\pi}(q^2) = f^+_{B\pi}(q^2) + \frac{q^2}{m_B^2-m_\pi^2} f^-(q^2) \,.
\label{eq:f0}
\ee 

Finally, the LCSR for the penguin form factor 
obtained from the third dispersion relation in (\ref{eq:disp})
has the following expression:
\ba
f^T_{B\pi}(q^2)= 
\frac{(m_B+m_\pi)e^{m_B^2/M^2}}{2m_B^2 f_B} 
\Bigg[F^T_{0}(q^2,M^2,s_0^B)+
\frac{\alpha_s C_F}{4\pi} F^T_{1}(q^2,M^2,s_0^B)\Bigg]\,,
\label{eq:fTLCSR}
\ea
where 
\ba
&& F_{0}^{T}(q^2,M^2,s_0^B)= m_b f_\pi\int\limits_{u_0}^1 
du\, e^{-\frac{m_b^2-q^2\bar{u}}{uM^2}} 
\Bigg\{\frac{\varphi_\pi(u)}{u}
-\frac{m_b\mu_\pi}{3(m_b^2-q^2)}
\frac{d\phi_{3\pi}^\sigma(u)}{du}
\nonumber
\\
&& 
+\frac{1}{m_b^2-q^2}
\Bigg(\frac{1}{4}\frac{d\phi_{4\pi}(u)}{du} -
\frac{m_b^2\,u}{2(m_b^2-q^2)}\frac{d^2\phi_{4\pi}(u)}{du^2}
-I_{4\pi}^T(u)\Bigg)\Bigg\}\,,
\label{eq:fTBpiLCSRcontrib}
\ea
and
\ba
&I_{4\pi}^T(u)&= \frac{d}{du}\Bigg(\int\limits_0^u\! d\alpha_1\!\!\!
\int\limits_{(u-\alpha_1)/(1-\alpha_1)}^1\!\!\!\!\! \frac{dv}{v} \,\,
\Bigg[2\Psi_{4\pi}(\alpha_i)-(1-2v)\Phi_{4\pi}(\alpha_i)
\nonumber
\\
&& \qquad +
2(1-2v)\widetilde{\Psi}_{4\pi}(\alpha_i)-
\widetilde{\Phi}_{4\pi}(\alpha_i)\Bigg]
\Bigg|_{\begin{array}{l}
\alpha_2=1-\alpha_1-\alpha_3,\\
\alpha_3=(u-\alpha_1)/v
\end{array}
}\Bigg)\,.
\label{eq:fTBpiLCSR3part}
\ea
The NLO parts $\widetilde{F}_1$ and $m_b F^T_1$  
in LCSR (\ref{eq:fplminLCSR}) and (\ref{eq:fTLCSR}),
respectively, are represented in the form similar to 
(\ref{eq:Imconvol}), and the corresponding 
imaginary parts are collected in App.~B.

For $f_B$ entering LCSR we use the well known two-point sum rule 
\cite{AE} obtained from the correlator of two 
$m_b\bar{q} i\gamma_5 b$ currents. The latest analyses of
this sum rule can be found in \cite{JL,PS}; here we 
employ the $\overline{MS}$  version \cite{JL}. 
For consistency with LCSR, the sum rule  
for $f_B$ is taken with $O(\alpha_s)$ accuracy. 
For convenience, this expression is written down in App.~C.

Note that the expressions for LCSR in LO are slightly modified 
as compared to the ones presented in the previous papers.
We prefer not to use the so-called ``surface terms'', which originate
from the powers of $1/(m_b^2-(q+up))^n$ with $n>1$  in the correlation
functions. Instead, we use a completely equivalent but 
more compact form, with derivatives of DA's.

The twist-2 NLO part of LCSR for $f^+_{B\pi}$,
hence, the expressions for   
$T_1$ and $\mbox{Im}T_1$ in App.~B, after transition
to the pole scheme (the additional expressions
necessary for this transition are also presented in App.~B) 
coincide with the ones obtained in \cite{KRWY}.
We have also checked an exact 
numerical coincidence with the twist-2 NLO part of the 
sum rule in \cite{BBB}, written in a different analytical form. 
The explicit expressions for the amplitudes $T_1^{p,\sigma}$,
$\widetilde{T}_1$, $\widetilde{T}_1^{p,\sigma}$,
and $T_1^T$,$T_1^{Tp,\sigma}$
and their imaginary parts presented in App.~B  
are new. The $O(\alpha_s)$ spectral density 
entering the LCSR for $f^+_{B\pi}$ is given in \cite{BZ04}
in  a different form, that is, with the $u$-integration performed,
making an analytical comparison of our result with 
this expression very complicated. The numerical comparison  
is discussed below, in sect.~6. 
Furthermore, in \cite{RWY} the LCSR for the form factor $f^0_{B\pi}$ 
was obtained,
and the imaginary part of $\widetilde{T}_1$ was presented. A 
comparison with our expression for $\mbox{Im} \widetilde{T}_1$
reveals, however, some differences.

Since the imaginary parts of the hard-scattering
amplitudes have a very cumbersome 
analytical structure, we carried out a special
check of these expressions. Each hard-scattering amplitude
$T_1,...$  taken as a function of $u,q^2,(p+q)^2$ 
was numerically compared with its dispersion relation in 
the variable $(p+q)^2=s$, 
where the  expression for $\mbox{Im}_s T_1,...$ was substituted. 
Note that one has to perform one subtraction in order
to render the dispersion integral convergent.
\FIGURE[t]{
\includegraphics[width=10cm]{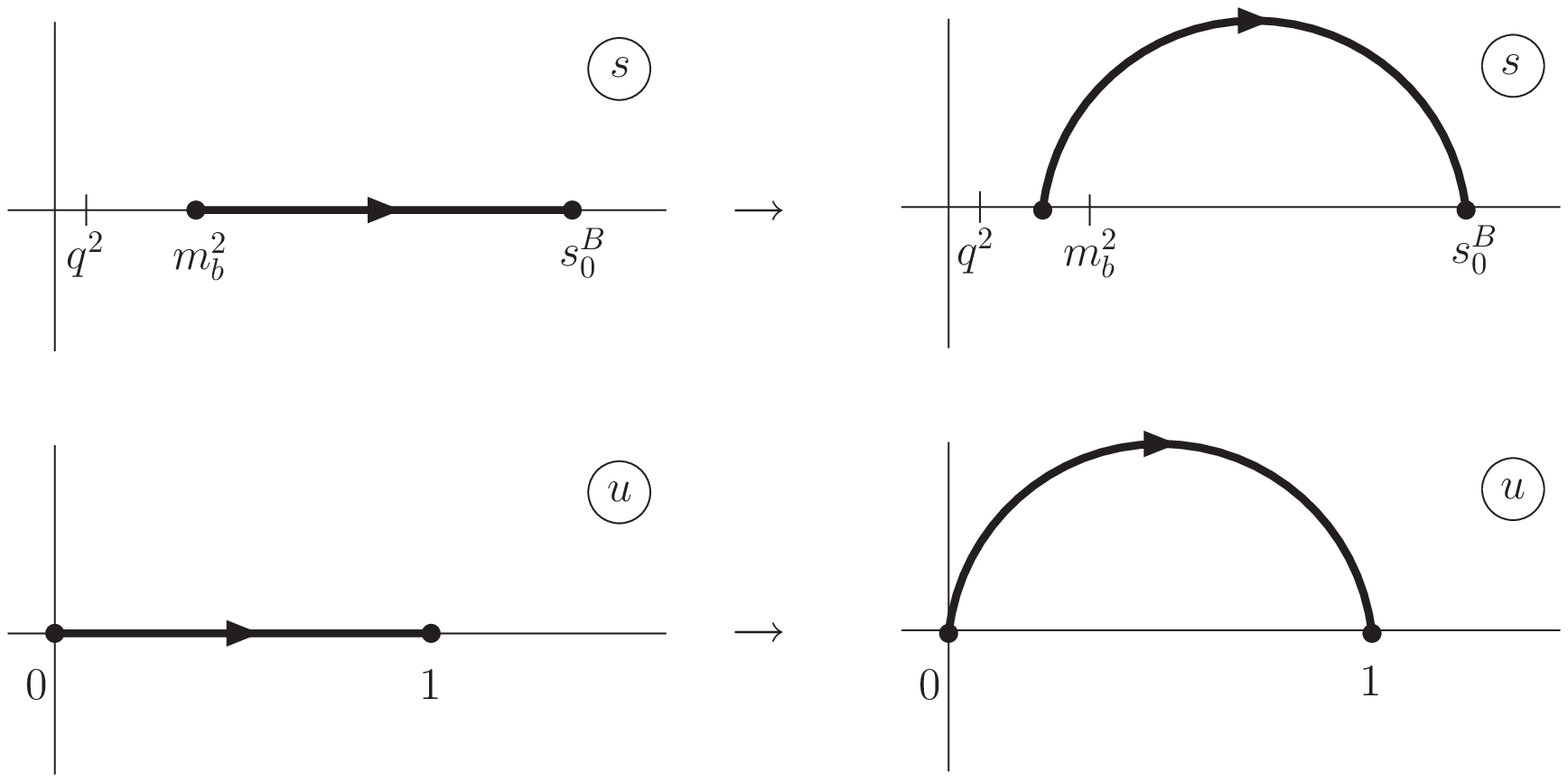}
\caption{\it Replacing the integration
intervals by the contours in the complex planes
of $u$ and $s$ variables in the alternative procedure
of the numerical integration of NLO amplitudes.
\label{fig:contour}}}

In addition, we applied a new  method which completely avoids  the use of
explicit imaginary parts of hard-scattering amplitudes,
allowing one to  numerically calculate the NLO parts of LCSR, e.g.,
$F_1(q^2,M^2,s_0^B)$ in (\ref{eq:Imconvol}),
analytically continuing integrals to the complex plane.
We make use of the fact that the hard-scattering amplitudes
$T_1,T_1^p,T_1^\sigma$ are analytical functions
of the variable $s=(p+q)^2$  in the upper half of the complex plane,
because of $i\epsilon$'s in Feynman propagators.
Consider, as an example the twist-2 part of $F_1$ given by
the integral over $s$ in the second line of  (\ref{eq:Imconvol}).
Since the integration is performed along the real axis, the operation of
taking the imaginary part can be moved outside the integral.
To proceed, one has to shift the lower limit
of the $s$-integration to any point
at $q^2<s<m_b^2$. This is legitimate because all $T_1$'s are real 
at $s<m_b^2$. Then one deforms the path of the $s$-integration,
replacing it by a contour in the upper half of the complex plane,
as shown schematically in Fig.~\ref{fig:contour},
so that all poles and cuts are away from the integration region.
Only when $s$ is approaching the upper limit $s_0^B$, 
one nears the pole at $u=(m_b^2-q^2)/(s_0^B-q^2)$
while performing the integration over $u$.
Because this pole does not touch the limits $u=0,1$,
it is possible to avoid it by moving the
contour of the $u$-integration into the upper half of the
complex $u$-plane (see Fig.~\ref{fig:contour}). After that, both
numerical integrations become completely stable.
Note, that in both $s$- and $u$-integrations, we integrate over the
semi-circle, but the
contour of the integration can be deformed in an
arbitrary way in the upper half of the complex plane.
The numerical integrations of $T_1$ over these contours
yield an imaginary part which represents the desired answer for $F_1$.
We have checked that the numerical
results obtained by this alternative method coincide with the ones obtained
by the direct integration over the imaginary parts,
thereby providing an independent check.

\section{Numerical results}
Let us specify the input parameters 
entering the LCSR (\ref{eq:fplusLCSR}), (\ref{eq:fplminLCSR}) 
and (\ref{eq:fTLCSR}) for $B\to \pi$ form factors 
and the two-point sum rule 
(\ref{eq:fBSRMSbar}) for $f_B$.

The value of the $b$-quark mass is taken  
from one of the most recent determinations \cite{KSS}:
\be
\overline{m}_b( \overline{m}_b)= 4.164 \pm 0.025 \; \mbox{GeV}\,,
\label{eq:bmass}
\ee
based on the bottomonium sum rules in the 
four-loop approximation. 
Note that (\ref{eq:bmass}) has a smaller uncertainty than the 
average over the non-lattice determinations given 
in \cite{PDG}: 
$
\overline{m}_b( \overline{m}_b)= 4.20 \pm 0.07 \; \mbox{GeV}\,.
$
However, as we shall see below, the uncertainty 
of $\overline{m}_b(\overline{m}_b)$ does not significantly influence
the ``error budget'' of the final prediction.
Furthermore, in our calculation, the scale-dependence 
$\overline{m}_b(\mu_m)$ is 
taken into account  in the one-loop approximation which is sufficient 
for the $O(\alpha_s)$-accuracy of the 
correlation function. Note that 
using the $\overline{MS}$  mass 
inevitably introduces some scale-dependence of the lower 
threshold $m_b^2$ in the dispersion integrals in both LCSR and 
$f_B$ sum rule. However, this does not create a problem, because 
the imaginary part of the OPE correlation
function obtained from a fixed-order perturbative QCD calculation
is not an observable, but only serves as an approximation
for the hadronic  spectral density.

The QCD coupling
$\alpha_s(\mu_r)$  is obtained from 
$\alpha_s(m_Z)= 0.1176 \pm 0.002\, ~$\cite{PDG}, with the NLO evolution 
to the renormalization scale $\mu_r$.
In addition to $\mu_m$ and $\mu_r$, one encounters
the factorization scale $\mu_f$ in the correlation
function, at which the pion DA's are taken. 
In what follows, we adopt a single scale 
$ \mu=\mu_m=\mu_r=\mu_f$
in both LCSR and two-point SR for $f_B$. The numerical value of $\mu$ 
will be specified below.

\TABLE[t]{
\begin{tabular}{|c|c|c|c|}
\hline
&&&\\
twist& Parameter & Value at $\mu=1 $ GeV& Source\\
&&&\\
\hline
2 & $a_2^\pi$ & $0.25 \pm 0.15$& average from \cite{BBL} \\
 & $a_4^\pi$ & $-a_2^\pi+(0.1\pm 0.1)$ &  $\pi\gamma\gamma^*$ form factor 
\cite{Bakulev}\\
 & $a_{> 4}^\pi$ & 0  & \\
\hline
 & $\mu_\pi$ & $ 1.74^{+0.67}_{-0.38}$ GeV& GMOR relation; $m_{u,d}$ from \cite{PDG}\\
3 & $f_{3\pi}$ & $0.0045\pm 0.0015$ GeV$^2$ &2-point QCD SR \cite{BBL}\\
 & $\omega_{3\pi}$& $ -1.5\pm 0.7$& 2-point QCD SR\cite{BBL}\\
\hline
4 & $\delta^2_{\pi}$& $0.18\pm 0.06$  GeV$^2$ &2-point QCD SR \cite{BBL}\\
& $\epsilon_{\pi}$&$ \frac{21}{8}(0.2\pm 0.1)
$& 2-point QCD SR \cite{BBL}\\
\hline
\end{tabular}
\\
\caption{\it Input parameters for the pion DA's.
\label{tab-1}}}

The input parameters of the twist-2 pion DA include 
$f_\pi=130.7 $ MeV  \cite{PDG} 
and  the two first Gegenbauer moments 
$a_2^\pi$ and $a_4^\pi$ normalized at a low scale 1 GeV.
For the latter we adopt the intervals presented in Table~\ref{tab-1}.
The range for $a_2^\pi(1 \mbox{GeV})$ is 
an average \cite{BBL} over various recent determinations, including, e.g.,  
$a_2^\pi(1 \mbox{GeV})=0.26^{+0.21}_{-0.09}$ 
calculated from the two-point sum rule  in \cite{KMM}. 
For $a_4^\pi $ we use, following \cite{BZ04}, 
the constraint $a_2^\pi(1 \mbox{GeV}) +a_4^\pi(1 \mbox{GeV})=0.1\pm 0.1$, 
obtained \cite{Bakulev} from the analysis of 
$\pi\gamma\gamma^*$ form factor.
Having in mind, that at large scales   
the renormalization suppresses all higher Gegenbauer 
moments, we set $a_{>4}^\pi=0$ 
in our ansatz for $\varphi_\pi(u)$ specified in App.~A. 
The uncertainties of $a_{2,4}^\pi(1 \mbox{GeV})$  remain large,
hence we neglect very small effects of 
their NLO evolution taken into account in \cite{KRWY}.

The normalization parameter $\mu_\pi(1 \mbox{GeV})$  
of the twist-3 two-particle DA's 
presented in Table~\ref{tab-1} is obtained 
adopting the (non-lattice) intervals \cite{PDG}
for the light quark masses: $m_u(2 ~\mbox{GeV})=3.0 \pm 1.0 ~\mbox{MeV}$, 
$m_d(2~\mbox{GeV})=6.0 \pm 1.5 ~\mbox{MeV}$.
Correspondingly, the quark-condensate density given by GMOR 
relation is: 
\be
\langle \bar{q}q \rangle (1 \mbox{GeV})=
-\frac12 f_\pi^2\mu_\pi(1 \mbox{GeV})=
-(246 ^{+ 28}_{-19} ~\mbox{MeV})^3\,,
\label{eq:cond}
\ee
where very small $O(m_{u,d}^2)$ corrections are neglected.
We prefer to use the above range, rather
than a narrower  ``standard'' interval 
$\langle \bar{q}q \rangle (1 \mbox{GeV})= -(240\pm 10 ~\mbox{MeV})^3$ 
employed in the previous analyses.
In fact, (\ref{eq:cond}) is consistent with 
$\langle \bar{q}q \rangle (1 \mbox{GeV})= (254\pm 8 ~\mbox{MeV})^3$  
quoted in the review \cite{Ioffe}, as well as with 
the recent determination of the light-quark masses from 
QCD sum rules with $O(\alpha_s^4)$ accuracy \cite{JOP}:
$m_u(2~\mbox{GeV}) = 2.7 \pm 0.4 ~\mbox{MeV}$,
$m_d(2~\mbox{GeV}) = 4.8 \pm 0.5 ~\mbox{MeV}$.

The remaining parameters of the twist-3 DA's ($ f_{3\pi}$,
$\omega_{3\pi}$) and twist-4 DA's ($\delta_\pi^2$, $\epsilon_\pi$)  
presented in Table~\ref{tab-1} are taken from \cite{BBL}, where 
they are calculated from auxiliary two-point sum rules. The latter  
are obtained from the  vacuum correlation functions 
containing the local quark-gluon operators that 
enter the matrix elements (\ref{eq:tw3matr}), (\ref{eq:tw3matr1}) 
and (\ref{eq:delta1}),
(\ref{eq:eps}).  
The one-loop running for all parameters of DA's is taken into account
using the scale-dependence relations presented in App.~A.
Note that the small value of $f_{3\pi}$  
effectively suppresses all nonasymptotic 
and three-particle contributions of the twist-3 DA's. 
Furthermore, the overall size of the twist-4 
contributions to LCSR 
is very small. Hence, although the parameters 
of the twist-3,4 DA's have large uncertainties,
only the accuracy of $\mu_\pi$ plays a role  in LCSR\footnote{We also expect that 
the use of the recently developed 
renormalon model \cite{renorm} for the twist-4 DA's, 
instead of the ``conventional'' twist-4 DA's 
\cite{BF} used here, will not noticeably change the numerical results.}.
Finally, in the sum rule (\ref{eq:fBSRMSbar}) for $f_B$ 
the gluon condensate density 
$\langle \alpha_s/\pi GG\rangle =0.012^{+0.006}_{-0.012}$ GeV$^4$
and the ratio of the quark-gluon and quark-condensate densities
$m_0^2=0.8 \pm 0.2 ~ \mbox{GeV} ^2$ \cite{Ioffe} are used, the 
accuracy of these parameters playing a minor role.

The universal parameters listed above determine 
the ``external'' input for sum rules. The next step 
is to specify appropriate intervals 
for the ``internal'' parameters:  
the scale $\mu$, the Borel parameters $M$ and $\overline{M}$
and the effective thresholds $s_0^B$ and $\overline{s}_0^B$.
In doing that, we take  all external input parameters 
at their central values, allowing only $a_2^\pi$ and $a_4^\pi$ to vary
within the intervals given in Table~\ref{tab-1}.  

From previous studies \cite{KRWY,BBB,KRWWY,BZ04} it is known that 
an optimal renormalization scale is 
$\mu \sim \sqrt{m_B^2-m_b^2}\sim \sqrt{2m_b\bar{\Lambda}}$ 
(where $\bar{\Lambda}$ does not scale with the heavy quark mass),
and simultaneously, $\mu$ has the order of magnitude of  the Borel scales 
defining the average virtuality in  the correlation functions.
In practice, $M$ and $\overline{M}$ 
are varied within the ``working windows'' of the respective sum rules, hence
one expects that also $\mu$ has to be taken in a certain interval.   

Calculating  the total  Borel-transformed correlation function 
(that is, the $s_0^B\to \infty$ limit of LCSR) 
we demand that the contribution of subleading twist-4 terms 
remains very small, $< 3\%$ of the LO twist-2 term,
thereby diminishing the contributions of the 
higher twists, that are not taken into account in the OPE. 
This condition puts a lower bound $M^2\geq M_{min}^2= 15$ GeV$^2$. 
In addition, in order to keep the $\alpha_s$-expansion 
in the Borel-transformed correlation function 
under control, both NLO  
twist-2 and twist-3  terms are kept $\leq 30\%$
of their LO counterparts,
yielding  a lower limit $\mu\geq 2.5 $ GeV. 
Hereafter a ``default'' value $\mu=3$ GeV is used.

Furthermore, we determine the effective threshold parameter 
$s_0^B$ in LCSR for each $M^2\geq M^2_{min}$. We refrain from using equal 
threshold parameters in LCSR and two-point sum rule for $f_B$, 
as it was done earlier, e.g. in \cite{KRWY,KRWWY}. 
Instead, we control the duality approximation
by calculating certain observables directly from LCSR and fitting them to 
their measured values.
Importantly, we include in the fitting procedure 
not only $s_0^B$, but also the two least restricted external 
parameters $a_2^\pi$ and $a_4^\pi$, under the condition that 
both Gegenbauer moments  
remain within the intervals of their direct determination 
given in Table~\ref{tab-1}. 

The first observable used in this analysis is the $B$-meson mass.
In a similar way, as e.g., in \cite{BZ04,JL},
$m_B^2$ is calculated taking the derivative of LCSR 
over $-1/M^2$ and dividing it by the original sum rule. The  
$B$-meson mass extracted from LCSR has to deviate from 
its experimental value $m_B=5.279 $ GeV by less than 1 \%. 
Secondly, we make use of the recent rather accurate measurement of the  
$q^2$-distribution in $B\to \pi l \nu$
by BABAR collaboration \cite{babarslope}. 
We remind that LCSR for $B\to \pi$ form factors are valid 
up to momentum transfers  $q^2\sim m_b^2-2m_b\bar{\Lambda}$ , typically
at $0<q^2< 14-15$  GeV$^2$. 
To be on the safe side, we take the maximal allowed $q^2$  slightly lower than 
in the previous analyses and calculate the slope  
$f^{+}_{B\pi}(q^2)/f^{+}_{B\pi}(0)$ from LCSR  at  $0<q^2<12$ GeV$^2$. 
The obtained ratio is then fitted to the slope of the form factor 
inferred from the data. We employ the result 
of \cite{BallVub}, where various parameterizations of the 
form factor $f^+_{B\pi}(q^2)$ are fitted to the
measured $q^2$-distribution. Since all fits turn out to 
be almost equally good, we adopt the simplest 
BK-parameterization \cite{BK}: 
\be
\frac{f^{+(BK)}_{B\pi}(q^2)}{f^{+(BK)}_{B\pi}(0)}= 
\frac{1}{(1-q^2/m_{B^*}^2)(1-\alpha_{BK}q^2/m_B^2)}
\label{eq:BKfit}
\ee
with  the slope parameter $\alpha_{BK}=0.53\pm 0.06$ from \cite{BallVub}  
(close to $\alpha_{BK}$ fitted in \cite{babarslope}).
\FIGURE[t]{
\includegraphics[width=10cm]{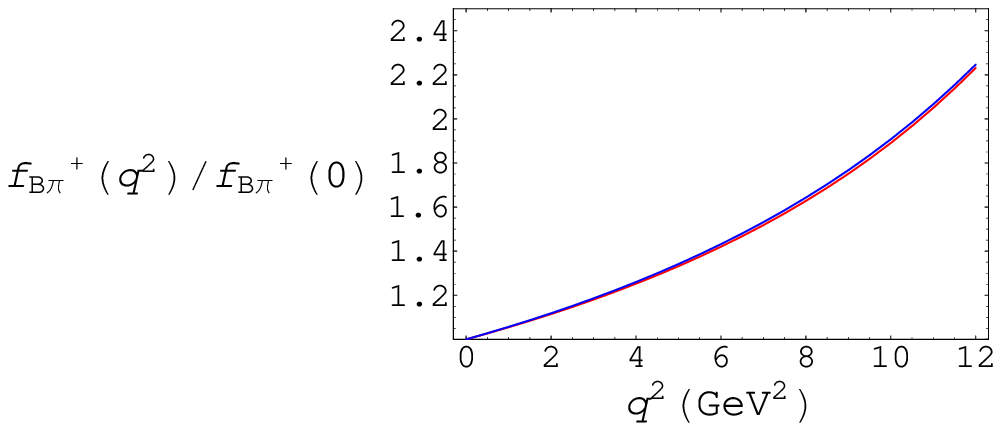}
\caption{\it The LCSR prediction for the 
form factor shape
$f^+_{B\pi}(q^2)/f^+_{B\pi}(0)$ 
fitted  to the BK parameterization 
of the measured $q^2$-distribution. The two (almost indistinguishable) 
curves are the 
fit and the parameterization (\ref{eq:BKfit}) at 
$\alpha_{BK}=0.53$. }
\label{fig-fit}
}

After fixing $s_0^B$ for each accessible $M^2$, we demand 
that heavier hadronic states contribute less than
30\% of the ground-state $B$ meson contribution
to LCSR. This condition yields  an upper limit $M^2<M_{max}^2=21$ GeV$^2$.
The resulting spread of the threshold parameter 
and Gegenbauer moments when $M^2$ varies between $M^2_{min}$ and 
$M^2_{max}$  
is very small: $s_0^B=36-35.5 $ GeV$^2$, 
$a_2^\pi(1 \mbox{GeV})=0.15-0.17$, $a_4^\pi(1 \mbox{GeV})=0.05-0.03$.
The quality of the fit is illustrated 
in Fig.~\ref{fig-fit} where the two curves: the calculated 
$q^2$-shape of the form factor 
and the BK-parameterization (\ref{eq:BKfit})  
are almost indistinguishable. Thus, in our numerical analysis 
we ``trade'' the $q^2$-dependence predicted from LCSR for a smaller 
uncertainty of the Gegenbauer moments and for a better 
control over the quark-hadron duality approximation. 

In the final stage of the numerical analysis we turn to 
the two-point QCD sum rule for $f_B$ presented in App.~C and 
find that at the adopted value 
of the renormalization scale  $\mu= 3 $ GeV  
the interval  $\overline{M}^2=5.0\pm 1.0$ GeV$^2$
satisfies the same criteria as the ones imposed in the numerical analysis of 
LCSR: the smallness of higher power terms in OPE,
and suppression of the heavier hadronic contributions. 
The threshold parameter $\overline{s}_0^B= 35.6^{-0.9}_{+2.1} $ GeV$^2$ 
is fixed  by calculating $m_B^2$ from this sum rule. This time the deviation 
from the experimental value is even less than 0.5 \%.
For completeness, we quote the resulting interval  
$f_B=214^{-5}_{+7}$ MeV.
Note that the $O(\alpha_s^2)$ correction taken into account in \cite{JL}
is not included here. 
As usual, employing the sum rule for $f_B$ in order to extract 
the form factor from the LCSR for the product $f_Bf_{B\pi}^+$ 
turns out to be extremely useful. One observes a partial cancellation 
of the $\alpha_s$-corrections in both LCSR 
and two-point sum rule and a better stability with respect to 
the variation of scales. 
\begin{figure}[t]
\begin{center}
\includegraphics[width=7cm]{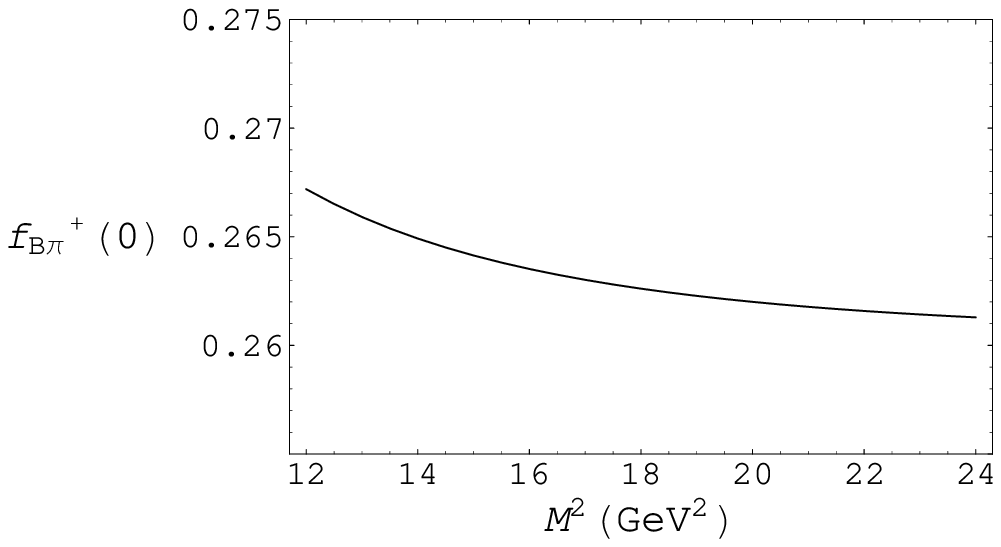}\hspace{1cm}
\includegraphics[width=7cm]{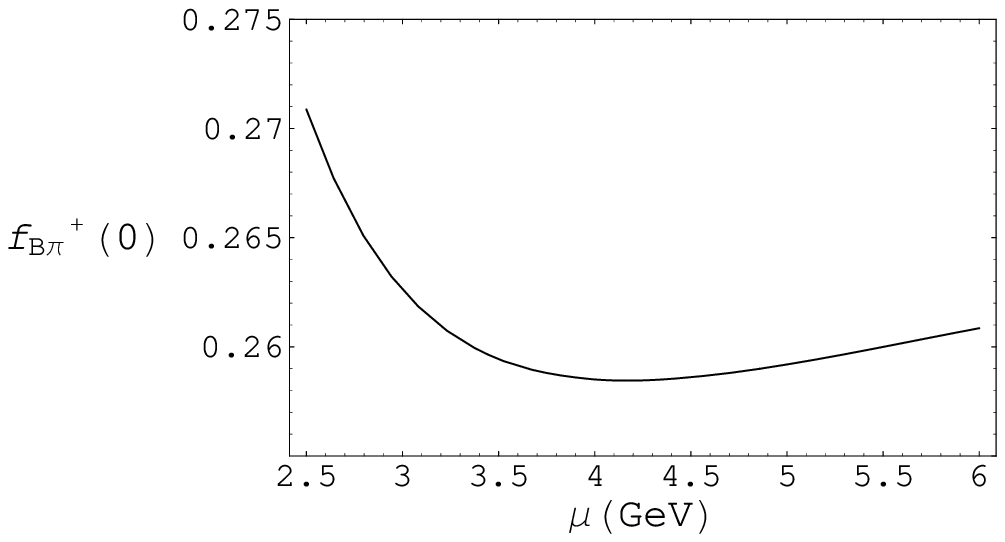}\\
\caption{ \it Dependence of $f^+_{B\pi}(0)$
on the Borel parameter (left) and 
renormalization scale (right).}
\label{fig-Mmudep}
\end{center}
\end{figure}  
\FIGURE[t]{
\includegraphics[width=10cm]{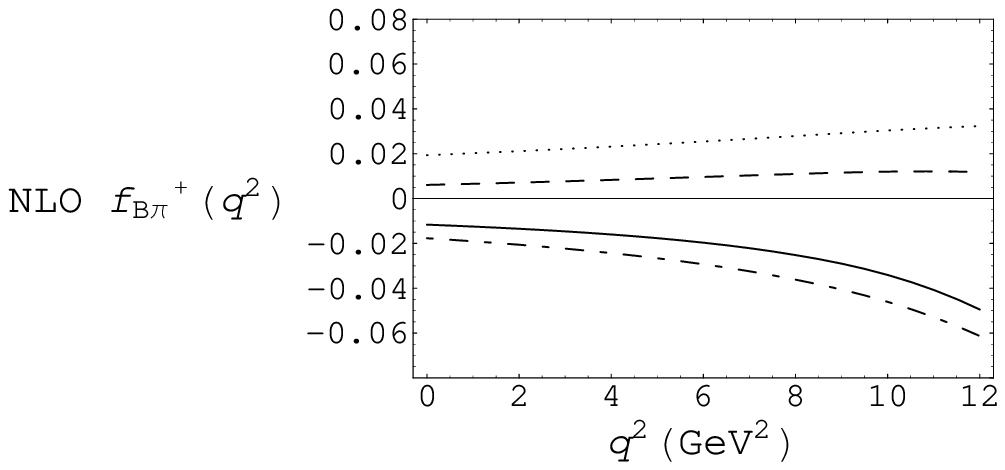}
\caption{ \it Gluon radiative corrections to the twist-2 
(dotted line) and twist-3 (solid line) parts of LCSR
for $f^+_{B\pi}(q^2)$, as a function of $q^2$.  
The part proportional to $\phi_\sigma$  ($\phi_p$) is shown 
separately by dashed (dash-dotted) line.}
\label{fig-nlo}
}

To demonstrate some important numerical features
of the LCSR prediction,  
in Fig.~\ref{fig-Mmudep} (left) we plot the $M^2$-dependence of the form factor 
$f^+_{B\pi}(0)$ with all other inputs fixed at their central values. 
The observed stability, far beyond the adopted ``working'' interval in 
$M^2$, serves as  
a usual criterion of reliability in QCD sum rule approach. 
The $\mu$-dependence plotted in Fig.~\ref{fig-Mmudep}~(right)
is very mild from $\mu=2.5$ GeV up to $\mu=6$ GeV. 
The numerical size of the gluon radiative corrections  
in LCSR is illustrated  in Fig.~\ref{fig-nlo}.

The numerical analysis yields the following   
prediction for the vector $B\to\pi$ form factor at zero momentum transfer:
\be
f^+_{B\pi}(0)= 0.263 ^{~+ 0.004}_{~-0.005}\bigg|_{M,\overline{M}}
\,_{-0.004}^{+0.009}\bigg|_{\mu} \pm 0.02 \bigg|_{shape}
\,^{+0.03}_{-0.02}\bigg|_{\mu_\pi}\pm 0.001\bigg|_{m_b}\,,   
\label{eq:fBpires}
\ee
where the central value is calculated at 
$\mu=3.0$ GeV, $M^2=18.0$ GeV$^2$, $s_0^B=35.75$ GeV$^2$,     
$a_2^\pi(1 \mbox{GeV})=0.16$, $a_4^\pi(1 \mbox{GeV}) = 0.04$,
$\overline{M}^2=5.0$ GeV$^2$ and $\overline{s}_0^B=35.6$ GeV$^2$.
The percentages of different contributions 
to the central value in (\ref{eq:fBpires}) are 
presented in Table~\ref{tab-2}.
\TABLE[b]{
\caption{\it The form factor $f^+_{B\pi}$ at zero momentum transfer 
calculated from LCSR in two different quark-mass schemes
and separate contributions to the sum rule in \%. 
}
\begin{tabular}{|c|c|c|}
\hline
$b$-quark mass & $\overline{MS}$& pole\\ 
\hline
input & central & set II from \cite{BZ04} \\
\hline
$f^+_{B\pi}(0)$ & 0.263 & 0.258\\\hline
tw2 LO & 50.5\%   & 39.7\% \\
tw2 NLO &  7.4\%  & 17.2 \% \\
\hline
tw3 LO & 46.7\% & 41.5 \% \\
tw3 NLO &-4.4\%  & 2.4 \% \\
\hline
tw4 LO &-0.2\%  & -0.9\% \\
\hline
\end{tabular}
\label{tab-2}
}

In (\ref{eq:fBpires}) the first (second) uncertainties are due to  
the variation of the Borel parameters  $M$ and $\overline{M}$ 
(scale $\mu$) 
within the intervals specified above. The third uncertainty
reflects the error of the experimental slope parameter.
In addition, we quote the uncertainties due to limited knowledge of 
the ``external'' input parameters. We have estimated them by 
simply varying these parameters one by one within their intervals
and fixing the central values for all ``internal'' input 
parameters. Interestingly, the largest 
uncertainty of order of 10\% is due to the error in 
the determination of light-quark masses transformed 
into the uncertainty of $\mu_\pi$, the coefficient of the large 
twist-3 LO contribution. The spread caused by  the 
current uncertainty of $\overline{m}_b(\overline{m}_b)$ 
is much smaller, hence, does not influence the resulting 
total uncertainty, even if one increases the error 
of $m_b$-determination by a factor of two. Remaining theoretical errors 
caused by the current uncertainties of $\alpha_s$, 
twist-3,4 DA's parameters and higher-dimensional condensates are very small,
and for brevity they are not shown in (\ref{eq:fBpires}).

Finally, we add all uncertainties in quadrature and obtain   
the interval: 
\be
f^+_{B\pi}(0)
= 0.26^{+0.04}_{-0.03}\,,
\label{eq:fplBpiresult}
\ee
which is our main numerical result. 
It can be used to normalize the experimentally measured shape, e.g., 
the one in (\ref{eq:BKfit}), yielding the form factor $f^+_{B\pi}(q^2)$ in 
the whole  $q^2$-range of $B\to \pi l \nu_l$.

With this prediction at hand, we are in a position 
to extract $|V_{ub}|$.
For that we use the interval 
\be
|V_{ub}|f^+_{B\pi}(0)=\bigg(9.1\pm 0.6\big|_{shape}\pm 0.3\big|_{BR}\bigg)\times 10^{-4}\,,
\label{eq:fvub}
\ee
inferred \cite{BallVub} from the 
measured $q^2$-shape \cite{babarslope} and average 
branching fraction of $B\to\pi l \nu_l$ \cite{HFAG}.
We obtain: 
\be
|V_{ub}|=\bigg(3.5 \pm 0.4\big|_{th} 
\pm 0.2\big|_{shape}\pm 0.1\big|_{BR}\bigg)\times 10^{-3}\,,
\label{eq:Vub}
\ee
where the first error is due to the estimated uncertainty 
of $f^+_{B\pi}(0)$ in (\ref{eq:fplBpiresult}), and the two remaining 
errors  originate from the experimental errors in (\ref{eq:fvub}). 
A possible small correlation between the shape
uncertainty of our prediction for the form factor 
and the experimental shape uncertainty is not taken into account.

The remaining two $B\to \pi$ form factors can now be predicted without 
any additional input. In particular, we adopt 
the same Borel parameter $M^2$ and effective threshold $s_0^B$, 
assuming that they only depend on the quantum numbers of the 
interpolating current for $B$ meson.
The scalar form factor 
$f^0_{B\pi}(q^2)$, obtained by combining the LCSR for $f^+_{B\pi}$  
and $(f^+_{B\pi}+f^-_{B\pi})$, and the penguin form factor 
$f^T_{B\pi}(q^2)$ 
are presented in Fig.~\ref{fig-ff}, in comparison with $f^+_{B\pi}(q^2)$. 
The predicted interval for the penguin form factor at zero momentum
transfer is :
\be
 f^T_{B\pi}(0)=0.255 \pm 0.035\,,
\label{eq:fTBpiresult}
\ee
adopting $\mu=3$ GeV as the  renormalization scale of the penguin current. 
\FIGURE[h]{
\includegraphics[width=10cm]{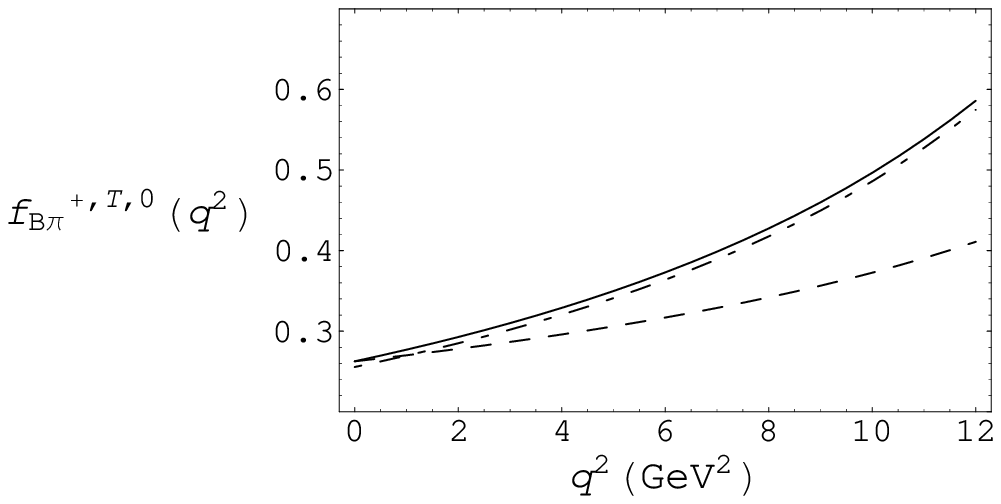}
\caption{ \it The LCSR prediction for 
form factors $f^+_{B\pi}(q^2)$ (solid line), $f^0_{B\pi}(q^2)$ (dashed line) 
and $f^T_{B\pi}(q^2)$ (dash-dotted line) 
at $0<q^2<12$ GeV$^2$ and for the central values of all input parameters. }
\label{fig-ff}
}

\section{Discussion}

In this paper, we returned to the LCSR for the 
$B\to \pi$ form factors.
We recalculated the 
$O(\alpha_s)$ gluon radiative corrections 
to the twist-2 and twist-3 hard-scattering amplitudes
and presented the first complete set of expressions for these 
amplitudes and their imaginary parts. 
For the radiative corrections to the twist-2 part of the LCSR  
for $f_{B\pi}^+$ we reproduced the results of \cite{KRWY,BBB}. 
For the radiative corrections to the twist-3 part we 
confirmed the cancellation of 
infrared divergences observed in \cite{BZ01,BZ04} 
in the case of asymptotic DA's. 
Including the nonasymptotic effects 
in these radiative corrections  
demands taking into account the mixing between two- 
and three-particle DA's. In fact, the 
parameter $f_{3\pi}$ determining 
the size of nonasymptotic twist-3 corrections is numerically small, 
hence these corrections are not expected  
to influence the numerical results.  

Throughout our calculation and in the final sum rule relations
we used the $\overline{MS}$-mass of the $b$ quark,
which is the most suitable mass definition for short-distance 
hard-scattering amplitudes.
Indeed, as follows from our numerical analysis,
the $O(\alpha_s)$ corrections to the sum rules turn out to be 
comparably small. To demonstrate that, we 
returned to the pole-mass scheme in LCSR
and used exactly the same input as in \cite{BZ04} 
(the preferred ``set 2'' with $m_b^{pole}=4.8\, {\rm GeV}$). 
We calculated the total form factor and separate contributions to the sum rule 
in both quark-mass schemes and compared them in Table~2. Note that   
the twist-2 NLO correction is distinctively smaller 
in the $\overline{MS}$-mass scheme. In the twist-3 part of the sum rule 
the $\alpha_s$-correction is small in both schemes.  
In the $\overline{MS}$ scheme, as seen from Fig.~\ref{fig-nlo},
this correction is dominated by the contribution of the DA $\phi^{p}_{3\pi}$.
In the pole scheme there is a partial cancellation 
between the contributions of the two twist-3 DA's.
Our numerical results for the  
form factors $f_{B\pi}^{+,0,T}(q^2)$ in the pole scheme are very close 
to the ones obtained in \cite{BZ04}. It is however difficult to
compare separate contributions, because they are not presented  
in \cite{BZ04}. We also cannot 
confirm the numerical values of the twist-3 NLO corrections 
to the form factor $f_{B\pi}^+(q^2)$ plotted in the figure presented in 
the earlier publication \cite{BZ01}.

Further improvements of  LCSR are possible but demand 
substantial calculational efforts. For example, obtaining 
radiative corrections to the three-particle twist-3,4
contributions is technically very difficult. Again, we expect 
no visible change of the predicted form factors 
because three-particle terms  are already very small in LO.
A more feasible task is to go beyond twist-4
in OPE and estimate the twist-5,6 effects, related
to the four-particle pion DA's, at least in the factorization approximation,
where one light quark-antiquark pair is replaced by the quark condensate
(In LCSR for the pion electromagnetic form factor these estimates  
have been done in \cite{BKM}).
  
The numerical analysis of LCSR was  improved 
due to the use of the $q^2$-shape measurement
in $B\to \pi l \nu$. A smaller theoretical uncertainty
of LCSR predictions can be anticipated 
with additional data on this shape, as well as with 
more accurate determinations of $b$- and, especially, $u,d$-quark masses.

In this paper, all calculations have been done in full QCD 
with a finite $b$-quark mass. At the same time, the whole
approach naturally relies on the fact that $m_b$ is a very large scale 
as compared with $\Lambda_{QCD}$ and related nonperturbative
parameters. Our results demonstrate that the twist-hierarchy 
as well as the perturbative expansion of the correlation function 
work reasonably well. An interesting problem is the investigation 
of the $m_b\to \infty$ 
limit of LCSR and various aspects of this limiting transition,
e.g., the hierarchy of radiative and nonasymptotic corrections. 
This problem remaining out of our scope 
was already discussed in several papers: earlier, 
in \cite{KRW} at the LO level, in \cite{BBB},\cite{BallSCET} at NLO level 
and more recently, in \cite{DFFH}.

\TABLE{
\caption{\it Recent $|V_{ub}|$ determinations  
from $B\to \pi l \nu_l$  }
\begin{tabular}{|c|c|c|c|}
\hline
[ref.] & $f^+_{B\pi}(q^2)$ calculation & $f^+_{B\pi}(q^2)$ input & 
 $|V_{ub}| \times 10^3$ \\ 
\hline
\cite{Okamoto}& lattice ($n_f=3$)  & - &  3.78$\pm$0.25$\pm$0.52\\
\hline
\cite{HPQCD}&  lattice ($n_f=3$) & - &  3.55$\pm$0.25$\pm0.50$ \\
\hline
\cite{Arnesen}&- & lattice $\oplus$ SCET $B\to \pi\pi$& 
$3.54\pm 0.17\pm 0.44$ \\
\hline
 \cite{BecherHill}&- & lattice &  $3.7\pm 0.2\pm 0.1$ \\
\hline
\cite{FlynnN} &- & lattice $\oplus$ LCSR &  $3.47\pm 0.29\pm 0.03$\\
\hline
\cite{BZ04,BallVub} & LCSR & - & $3.5\pm 0.4\pm 0.1$ \\
\hline
this work& LCSR   & -&   $ 3.5\pm 0.4 \pm 0.2\pm 0.1$ \\
\hline
\end{tabular}
\label{tab-3}
}
Finally, in Table~\ref{tab-3} we compare our result 
for $|V_{ub}|$ with the one of the previous LCSR analysis
and with the recent lattice QCD determinations obtained at 
large $q^2$ and extrapolated to small $q^2$ with the help of 
various parameterizations. The observed mutual agreement ensures 
confidence in the 
continuously  improving $V_{ub}$ determination from exclusive $B$ decays.

\section*{Acknowledgments}
We are grateful to Th. Feldmann, M.~Jamin and R.~Zwicky  
for useful discussions. This work was supported by the 
Deutsche Forschungsgemeinschaft 
(Project KH 205/1-2). 
The work of G.D and B.M was supported by the
Ministry of Science, Education and Sport of the 
Republic of Croatia, under contract 098-0982930-2864.
The partial support of A.~von Humboldt Foundation 
under the Program of Institute Partnership 
is acknowledged. The work of N.O. was supported  
by FLAVIAnet (Contract No. MRTN-CT-2006-035482).

\appendix

\section{Pion distribution amplitudes }

For convenience, we specify the set of 
the pion DA's and their parameters used in this paper. 
The notations and parameters for twist-3 and 4 DA's 
are taken from \cite{BBL},
where the earlier studies \cite{BF,BallDA} are updated.

The two-particle DA's of the pion  
enter the following decomposition of the 
bilocal vacuum-pion matrix element 
(for definiteness, $\pi^+$ in the final state):
\ba
&& \langle \pi^+(p)|\bar{u}_\omega^i(x_1) 
d^j_\xi(x_2)|0\rangle_{x^2\to 0}
  = \frac{i\delta^{ij}}{12}f_\pi 
\int_0^1 du~e^{iu p\cdot x_1 +i\bar{u}p\cdot x_2}
\Bigg ( [\DS p \gamma_5]_{\xi\omega} \varphi_\pi(u)
\nonumber\\
&&\qquad\qquad -[\gamma_5]_{\xi\omega}\mu_\pi\phi^p_{3\pi}(u)
+\frac 16[\sigma_{\beta\tau}\gamma_5]_{\xi\omega}p_\beta(x_1-x_2)_\tau \mu_\pi\phi^\sigma_{3\pi}(u)
\nonumber\\
&&\qquad\qquad  +\frac1{16}[\DS p \gamma_5]_{\xi\omega}(x_1-x_2)^2\phi_{4\pi}(u)
-\frac{i}{2} [(\DS x_1-\DS x_2) \gamma_5]_{\xi\omega}
\int\limits_0^u\psi_{4\pi}(v)dv 
\Bigg)
\,,
\label{eq:phipi}
\ea
In the above, the product of the quark fields is 
expanded near the light-cone, that is, $x_i=\xi_ix$, where $\xi_i$ 
are arbitrary numbers, and $x^2=0$; $\bar{u}=1-u$. 
The path-ordered gauge-factor (Wilson line) is omitted assuming the 
fixed-point gauge  for the gluons. The light-cone expansion 
includes the twist-2 DA $\varphi_\pi$, two twist-3 DA's  
$\phi_{3\pi}^p$, $\phi_{3\pi}^\sigma$
and two twist-4 DA's  $\phi_{4\pi}$ and $\psi_{4\pi}$. 
The usual definitions of DA's are easily obtained,
multiplying both parts of (\ref{eq:phipi})
by the corresponding combinations of $\gamma$ matrices and taking
Dirac and color traces.  

The decomposition of the three-particle 
quark-antiquark-gluon matrix element is:
\ba
&& \langle \pi^+(p)|\bar{u}_\omega^i(x_1) g_sG_{\mu\nu}^a(x_3)
d^j_\xi(x_2)|0
\rangle_{x^2\to 0}
=\frac{\lambda^a_{ji}}{32}\int {\cal D}\alpha_ie^{ip(\alpha_1 x_1+\alpha_2 x_2+\alpha_3x_3)}
\nonumber
\\
&& \times\Bigg[if_{3\pi}(\sigma_{\lambda\rho}\gamma_5)_{\xi
\omega}(p_\mu p_\lambda g_{\nu\rho}-
p_\nu p_\lambda g_{\mu\rho})\Phi_{3\pi}(\alpha_i)
\nonumber
\\
&& -f_\pi(\gamma_\lambda\gamma_5)_{\xi\omega}\Big\{(p_\nu 
g_{\mu\lambda}-p_\mu g_{\nu\lambda})\Psi_{4\pi}(\alpha_i)
+\frac{p_\lambda(p_\mu x_\nu-p_\nu x_\mu)}{(p\cdot x)}
\left(\Phi_{4\pi}(\alpha_i)+\Psi_{4\pi}(\alpha_i)\right)
\Big\}
\nonumber
\\
&& -\frac{if_\pi}2\epsilon_{\mu\nu\delta\rho}(\gamma_\lambda)_{\xi\omega}
\Big\{(p^\rho g^{\delta\lambda}-p^\delta g^{\rho\lambda})
\widetilde{\Psi}_{4\pi}(\alpha_i)+  
\frac{p_\lambda(p^\delta x^\rho-p^\rho x^\delta)}{(p\cdot x)}
\left(\widetilde{\Phi}_{4\pi}(\alpha_i)+
\widetilde{\Psi}_{4\pi}(\alpha_i)\right)\Big\}\Bigg]\,.
\nonumber 
\\
\label{eq:decomp2}
\ea
including one twist-3 DA $\Phi_{3\pi}$  and four twist-4 
DA's :  $\Phi_{4\pi}$, $\Psi_{4\pi}$, $\widetilde{\Phi}_{4\pi}$ and 
$\widetilde{\Psi}_{4\pi}$.
Here the convention $\epsilon^{0123}=-1$ is used, which
corresponds to $Tr\{\gamma^5  \gamma^\mu\gamma^\nu\gamma^\alpha
\gamma^\beta\}=4i\epsilon^{\mu\nu\alpha\beta}$ .

The following expressions for the DA's entering the  
decompositions (\ref{eq:phipi}) and (\ref{eq:decomp2}) are used: 
\begin{itemize}

\item twist-2 DA:
\be
\varphi_\pi(u)=6u\bar{u}\Big(1 +a_2C_2^{3/2}(u-\bar{u}) 
+a_4C_4^{3/2}(u-\bar{u})\Big)\,, 
\label{eq:tw2DA}
\ee
where, according to our choice, the first two Gegenbauer polynomials are included 
in the nonasymptotic part, with the coefficients
having the following LO scale dependence:
\be 
a_2(\mu_2)= \left[L(\mu_2,\mu_1)\right]
^{\frac{25C_F}{6\beta_0}} a_2(\mu_1),~~
a_4(\mu_2)= \left[L(\mu_2,\mu_1)\right]
^{\frac{91C_F}{15\beta_0}} a_4(\mu_1)
\label{eq:a2scale}
\ee
with $L(\mu_2,\mu_1)= \alpha_s(\mu_2)/\alpha_s(\mu_1)$,
$\beta_0=11-2n_f/3$.

\item twist-3 DA's :
\be
\Phi_{3\pi}(\alpha_i)=360\alpha_1\alpha_2\alpha_3^2\left[1+
\frac{\omega_{3\pi}}{2}(7\alpha_3-3)\right]
\ee
with the nonperturbative parameters $f_{3\pi}$ and 
$\omega_{3\pi}$ defined via matrix
elements of the following local operators:
\ba 
\langle \pi^+(p) |\bar{u}
\sigma_{\mu\nu}\gamma_5 G_{\alpha\beta}d
| 0 \rangle 
=if_{3\pi}\Big[(p_\alpha p_\mu g_{\beta\nu}-p_\beta p_\mu g_{\alpha\nu})
-(p_\alpha p_\nu g_{\beta\mu}-p_\beta p_\nu g_{\alpha\mu})\Big],
\label{eq:tw3matr}
\ea
\be
\langle \pi^+(p) |\bar{u}\sigma_{\mu\lambda}\gamma_5
[D_\beta,G_{\alpha\lambda}] d-
\frac37 \partial_\beta\bar{u}\sigma_{\mu\lambda}\gamma_5
G_{\alpha\lambda}d
|0\rangle= -\frac{3}{14}f_{3\pi}\omega_{3\pi}p_\alpha p_\beta p_\mu\,. 
\label{eq:tw3matr1}
\ee

The scale dependence of the twist-3 parameters is given by:
\begin{eqnarray}
\mu_{\pi}(\mu_2) &=& \left[L(\mu_2,\mu_1)\right]
^{-\frac{4}{\beta_0}}\mu_{\pi}(\mu_1)\,,~~~
f_{3\pi}(\mu_2) = \left[L(\mu_2,\mu_1)\right]^
{\frac{1}{\beta_0}\left(\frac{7 C_F}{3}+3\right)}
f_{3\pi}(\mu_1)\,,
\\
&& (f_{3\pi}\omega_{3\pi})(\mu_2) = 
\left[L(\mu_2,\mu_1)\right]^{\frac{1}{\beta_0}\left(\frac{7 C_F}{6}+10\right)}
(f_{3\pi}\omega_{3\pi})(\mu_1)\,.
\end{eqnarray}

The corresponding expressions for the twist-3 quark-antiquark DA are:
\ba
\phi^p_{3\pi}(u)=1+30\frac{f_{3\pi}}{\mu_\pi f_\pi} C_2^{1/2}(u-\bar u)-
3\frac{f_{3\pi}\omega_{3\pi}}{\mu_\pi f_\pi} C_4^{1/2}(u-\bar{u}),
\nonumber
\\
\phi^\sigma_{3\pi}(u)=6u(1-u)\left(1+5\frac{f_{3\pi}}{\mu_\pi f_\pi}
\left(1-\frac{\omega_{3\pi}}{10}\right)C_2^{3/2}(u-\bar u)\right).
\ea

\item twist-4 DA's:
\ba
\Phi_{4\pi}(\alpha_i)&=&120\delta_\pi^2\varepsilon_\pi 
(\alpha_1-\alpha_2)\alpha_1\alpha_2\alpha_3~,
\nonumber\\
\Psi_{4\pi}(\alpha_i)&=&30\delta_\pi^2(\mu) 
(\alpha_1-\alpha_2)\alpha_3^2[\frac13+2
\varepsilon _\pi(1-2\alpha_3)] ~,
\nonumber
\\
\widetilde{\Phi}_{4\pi} (\alpha_i)&=&-120\delta_\pi^2\alpha_1\alpha_2\alpha_3[\frac13+
\varepsilon_\pi (1-3\alpha_3)] ~,
\nonumber
\\
\widetilde{\Psi}_{4\pi}(\alpha_i)&=&30\delta_\pi^2\alpha_3^2(1-\alpha_3)[\frac13+2
\varepsilon_\pi (1-2\alpha_3)] ~,
\label{eq:tw4gluon}
\ea
are the four three-particle DA's, 
where the nonperturbative parameters $\delta^2_\pi$ and $\epsilon_\pi$ are 
defined as
\begin{equation}
\langle \pi^+(p) |\bar{u}\widetilde{G}_{\alpha\mu}\gamma^\alpha d
|0 \rangle=i\delta^2_\pi f_\pi p_\mu \,,
\label{eq:delta1}
\end{equation}
and (up to twist 5 corrections):
\be
\langle \pi^+(p)  |\bar{u}[D_\mu,\widetilde{G}_{\nu\xi}]\gamma^\xi d-
\frac49\partial_\mu\bar{u}\widetilde{G}_{\nu\xi}\gamma^\xi d
|0\rangle= -\frac{8}{21}f_\pi\delta_\pi^2\epsilon_\pi p_\mu 
p_\nu \, ,
\label{eq:eps}
\ee
with the scale-dependence:
\be
\delta^2_\pi(\mu_2) = \left[L(\mu_2,\mu_1)\right]
^{\frac{8C_F}{3\beta_0}}\delta^2_\pi(\mu_1)\,,~~
(\delta_\pi^2\epsilon_\pi)(\mu_2) = 
\left[L(\mu_2,\mu_1)\right]
^{\frac{10}{\beta_0}}(\delta^2_\pi\epsilon_\pi)(\mu_1)\,.
\ee
Note that the twist-4 parameter $\omega_{4\pi}$ 
introduced in \cite{BBL} is replaced by 
$\epsilon_\pi=(21/8)\omega_{4\pi}$. 

Correspondingly, the two-particle DA's of twist 4 are:
\ba
\phi_{4\pi}(u) &=& \frac{200}3\delta_\pi^2 u^2\bar{u}^2+
8\delta_\pi^2\epsilon_{\pi}\Big\{u\bar{u}(2+13u\bar{u})+
2u^3(10 -15u + 6u^2)\ln u 
\nonumber \\
&& +2\bar{u}^3(10 -15\bar u + 6\bar u^2)\ln \bar u \Big \}\,,
\\
\psi_{4\pi}(u)&=&\frac{20}3\delta_\pi^2C_2^{1/2}(2u-1)\,.
\label{eq:tw42partDAs}
\ea

\end{itemize}

These DA's  are related to the original definitions \cite{BF} as 

\be
\phi_{4\pi}(u)= 16\Big(g_1(u)-\int\limits_0^u g_2(v)dv\Big),~~~ 
\psi_{4\pi}(u)=-2\frac{dg_2(u)}{du} \,. 
\ee

\section{Formulae for gluon radiative corrections}

Here we collect the expressions for the hard-scattering amplitudes 
entering the factorization formulae (\ref{eq:convol}) 
and the resulting imaginary parts of these amplitudes  
determining the radiative correction (\ref{eq:Imconvol})
to LCSR (\ref{eq:fplusLCSR}) for $f^+_{B\pi}$, as well as the analogous 
expressions for LCSR (\ref{eq:fplminLCSR}) and (\ref{eq:fTLCSR}) for 
the other two form factors.  

To compactify the formulae, we use the dimensionless variables 
\be
r_1=\frac{q^2}{m_b^2},~~r_2=\frac{(p+q)^2}{m_b^2},
\ee
(in the imaginary parts $r_2=s/m_b^2$) and the integration variable :
\be
\rho=r_1+u(r_2-r_1)\,~~~ 
\int\limits_0^1 du=\int\limits_{r_1}^{r_2}\frac{d\rho}{r_2-r_1}\,,
\ee
and introduce the combinations of logarithmic functions  
\be
G(x)= \mbox{Li}_2(x)+\ln^2(1-x)+
\ln(1-x)\left(\ln \frac{m_b^2}{\mu^2}-1 \right)\,,
\label{eq:Gfunct}
\ee
where $\mbox{Li}_2(x)=-\int_0^x\frac{dt}t \ln(1-t)$  is the Spence function, and
\be
L_1(x) = \log \left (\frac{(x-1)^2}{x}\frac{m_b^2}{\mu^2} \right ) - 1\,,
~~~
L_2(x) = \log \left (\frac{(x-1)^2}{x}\frac{m_b^2}{\mu^2} \right ) - \frac{1}{x}\,\,.
\ee

The imaginary parts of the hard-scattering amplitudes 
are taken at fixed $q^2<m_b^2$ ($r_1<1$), analytically continuing 
these amplitudes in the variable $s=(p+q)^2$ (or $r_2$). The result contains 
combinations of $\theta(1-\rho)$, $\theta(\rho-1)$ and $\delta(\rho-1)$ 
and its derivatives. To isolate the spurious infrared
divergences which one encounters by taking the imaginary part, 
we follow \cite{KRWY} and introduce the usual plus-prescription 
\ba
\int_{r_1}^{r_2} d\rho \, 
 \left ( \left \{ 
\begin{array}{c}
\theta (1 -\rho) \\
\theta (\rho-1) 
\end{array} \right \} 
\frac{g(\rho)}{\rho-1} \right )_+
\phi(\rho) 
= \int_{r_1}^{r_2} d \rho 
\left \{ 
\begin{array}{c}
\theta (1 -\rho) \\
\theta (\rho-1) 
\end{array} \right \} 
\frac{g(\rho)}{\rho-1}\,\Big ( \phi(\rho) - \phi(1) \Big ),
\label{eq:plusD}
\ea
for generic functions $\phi(\rho)$, $g(\rho)$. 
Furthermore, to make the formulae for imaginary parts 
more explicit, we partially integrate 
the derivatives of $\delta(\rho-1)$ using, e.g.: 
\be
\int_{r_1}^{r_2}\!\!d\rho\,\delta '(\rho-1)\phi(\rho)= 
\int d\rho \delta(\rho-1) \left( 
-\frac{d}{d\rho}+\delta(r_2-1)\right)\phi(\rho)\,,
\ee
omitting the terms with $\delta(r_2-1)$ in all cases where $\phi(1)=0$.

\subsection{Amplitudes for $f_{B\pi}^+$ LCSR  }
\ba
\frac{1}{2} T_1 &=& 
\left(\frac{1}{\rho -1} - \frac{\text{r2}-1}{(\text{r2}-\text{r1})^2 u}\right) G(\text{r1})
+\left(\frac{1}{\rho-1} + \frac{1-\text{r1}}{(\text{r2}-\text{r1})^2 (1-u)}\right) G(\text{r2})
\nonumber \\
&&-\left( \frac{2}{\rho-1} -\frac{\text{r2}-1}{(\text{r2}-\text{r1})^2 u} + \frac{1-\text{r1}}{(\text{r2}-\text{r1})^2 (1-u)}
\right) G(\rho )
\nonumber \\
& &  + \frac{1}{r_2} \left(
\frac{\text{r2}-1}{\rho-1} - \frac{\text{r2}-1}{(\text{r2}-\text{r1}) (1-u)}\right) 
\log (1-\text{r2})
\nonumber \\
&& 
+ \frac{1}{r_2} \left(\frac{\text{r2}-2}{2 \rho } - \frac{r_2}{2 \rho ^2} 
+\frac{\text{r2}-1}{(\text{r2}-\text{r1})(1-u)} \right) \log (1-\rho )
\nonumber \\
&&+\frac{\rho +1}{2 (\rho -1)^2} \left ( 3 \log \left(\frac{m_b^2}{\mu^2}\right) - \frac{3 \rho + 1}{\rho} \right )\, , 
\label{eq:t1}
\ea
\ba
&&-\frac{1}{2 \pi}{\rm Im}_s T_1 =
\theta(1-\rho )
\left [ 
\frac{ 1-\text{r1}}{(\text{r2}-\text{r1}) (\text{r2}-\rho )} L_1(r_2)
+ \left(\frac{L_2(r_2)}{\rho -1}\right)_+  + \frac{1}{(r_2-\rho)}\left (\frac{1}{r_2} -1 \right ) \right ]
\nonumber \\
&& + \theta (\rho -1)\left [
\frac{ 1-\text{r1}}{(\text{r2}-\text{r1}) (\text{r2}-\rho )} L_1(r_2)
+\frac{ 1 + \rho - \text{r1}-\text{r2}}{(\text{r1}-\rho ) (\text{r2}-\rho )} L_1(\rho) 
+ \left(\frac{L_2(r_2) - 2 L_1(\rho)}{\rho -1}\right )_+
\right . \nonumber \\
&& \left . 
\qquad +\frac{1}{2\rho}\left ( 1 - \frac{1}{\rho} - \frac{2}{r_2} \right )\right ]
\nonumber \\
&& +\delta(\rho -1) \left [ \left (\log \frac{r_2 -1}{1- r_1} \right )^2 -
\left (\frac{1}{r_2} -1 + \log r_2 \right ) \log \frac{(r_2-1)^2}{1-r_1}
+ \frac{1}{2} \left ( 4 -3 \log \left(\frac{m_b^2}{\mu ^2}\right) \right )
\right . \nonumber \\
&& \left . 
\qquad + {\rm Li}_2(\text{r1}) - 3 \, {\rm Li}_2(1-\text{r2})+1- \frac{\pi^2}{2} 
 - \left (4-3 \log \left(\frac{m_b^2}{\mu ^2}\right)\right ) \left(1 + \frac{d}{d\rho}\right)
\right ]\,,
\ea
%
%
%
\ba
\frac{r_2-r_1}{2}T_1^{p} &=& 
\left(\frac{1}{ \rho -1} - \frac{4 \text{r1}-1}{(\text{r2}-\text{r1}) u}
\right) G(\text{r1})
-\left(\frac{\text{r1}}{ \rho-1}+
\frac{1+ \text{r1}+\text{r2}}{ (\text{r2}-\text{r1}) (1-u)}\right) G(\text{r2})
\nonumber \\
&&+ \left(-\frac{1- \text{r1}}{ \rho-1}+\frac{ 1+\text{r1}+\text{r2}}{ (\text{r2}-\text{r1}) (1-u)}
+\frac{4 \text{r1}-1}{ (\text{r2}-\text{r1}) u}\right) G(\rho )
\nonumber \\
&&-\left(\frac{ \text{r1}}{\rho-1} + \frac{2 \text{r1}}{(\text{r2}-\text{r1}) u}
\right) \log (1-\text{r1})
\nonumber \\
&&+ \frac{1}{r_2} \left(\frac{ \text{r1} + \text{r2} - \text{r1} \text{r2}}{\rho-1}
+\frac{\text{r1} - \text{r2} - \text{r2} ( \text{r1}  + \text{r2} ) }{ (\text{r2}-\text{r1}) (1-u)}\right) 
\log (1-\text{r2})
\nonumber \\
&&+\frac{1}{2} \left(\frac{3(3 - \text{r1})}{ \rho-1} + 
\frac{6 (1-\text{r1})}{ (\rho-1)^2} -1 \right)
\log \left(\frac{m_b^2}{\mu ^2}\right)
\nonumber \\
&&+\left(\frac{1-\text{r1}}{ \rho-1}
-\frac{1}{2} 
-\frac{ \text{r1} - \text{r2} - \text{r2}( \text{r1} +\text{r2})}{ \text{r2} (\text{r2}-\text{r1}) (1-u)}
+\frac{2 \text{r1}}{ (\text{r2}-\text{r1}) u}
\right . 
\nonumber \\
&& \left . -\frac{2 \text{r1}+\text{r2} - 3 \text{r1} \text{r2}}{ 2 \text{r2} \rho }
-\frac{\text{r1}}{2 \rho ^2}\right) \log (1-\rho ) 
 +\frac{2 (\text{r1}-3)}{ \rho-1}
+\frac{1}{2}-\frac{\text{r1}}{ 2 \rho }-\frac{4 (1-\text{r1})}{ (\rho-1)^2}\,,
\nonumber \\
\label{eq:t1p}
\ea
\ba
&&\frac{r_2-r_1}{2 \pi}{\rm Im}_s T_1^{p} =
\theta (1-\rho )
\left[\frac{ 1+ r_2 }{r_2 (r_2 -\rho)}
+\frac{ 1+ \text{r1}+\text{r2} }{r_2-\rho } L_2(r_2) - 
\left (1- r_1 L_2(r_2) \right )
\left(\frac{1}{\rho -1}\right)_+ \right ]
\nonumber \\
&& +\theta (\rho -1)
\left [ \frac{1+ \text{r1}+\text{r2} }{\text{r2}-\rho} L_1(r_2)
-  \left(\frac{4 \text{r1}-1}{ \rho -\text{r1}}
-\frac{ 1+ \text{r1}+\text{r2}}{\rho -\text{r2}}\right) L_1(\rho)
\right . \nonumber \\
&& \left .
\qquad 
+ \Bigg(\frac{ r_1 L_2(r_2) + \left (1 - r_1 \right )L_1(\rho) + r_1 - 2}
{\rho -1}\Bigg)_+
+ \frac{1}{2}  + \frac{2 r_1 + r_2 - 3 r_1 r_2}{2 r_2 \rho} + \frac{r_1}{2 \rho^2} + \frac{2 r_1}{r_1 -\rho}
\right ]
\nonumber \\
&& +\delta(\rho -1)
\left [ - \left (\log \frac{r_2 -1}{1- r_1} \right )^2
+  (r_1+1) \log \left( \frac{r_2 -1}{1-r_1} \right ) L_1(r_2)
\right . \nonumber \\
&& \left . 
\qquad - \log (r_2 -1) \left ( 2 \frac{r_1}{r_2} + 3 ( 1 - r_1) + ( r_1-1) \log r_2  \right )
+  \log (1 - r_1) \left ( \frac{r_1}{r_2} + 1  -  \log r_2  \right )
\right . \nonumber \\
&&\left . 
\qquad -\frac{\pi^2}{6} (4 \text{r1}+1) + \frac{1}{2}(\text{r1}-3) \left (3 \log \left(\frac{m_b^2}{\mu ^2}\right) -  4 \right )
- {\rm Li}_2(\text{r1})+ (1 - 2 r_1){\rm Li}_2(1-\text{r2})
\right . \nonumber \\
&&\left . 
\qquad + (1-\text{r1})
\left(4 - 3 \log \left(\frac{m_b^2}{\mu ^2}\right)\right) 
\left ( \frac{d}{d\rho}  - \delta (r_2-1) \right )\right ]\,,
\ea
%
\ba
3 T_1^{\sigma} &=& 
\left(
-\frac{1}{(\rho-1)^2} + \frac{2 (1-2 \text{r1})}{(1-\text{r1}) (\rho-1)} - 
\frac{1-4 \text{r1}}{(\text{r2}-\text{r1})^2 u^2}-\frac{2 (1- 2 \text{r1})}{(1-\text{r1}) (\text{r2}-\text{r1}) u}
\right) G(\text{r1})
\nonumber \\
&+&\left(
-\frac{\text{r1}}{\text{ }(\rho-1 )^2}
+\frac{2 \text{r2}}{(\text{r2}-1) (\rho-1)}
+\frac{1+\text{r1}+\text{r2}}{\text{ }(\text{r2}-\text{r1})^2 (1-u)^2}
+\frac{2 \text{r2}}{ (\text{r2}-\text{r1}) (\text{r2}-1) (1-u)}
\right) G(\text{r2})
\nonumber \\
&+&\left( \frac{1+\text{r1}}{\text{  }(\rho-1)^2}
+\frac{2 (1 -2 \text{r1}-2 \text{r2} + 3 r_1 r_2)}{\text{ }(1-\text{r1}) (\text{r2}-1) (\rho-1)}
-\frac{1+ \text{r1}+\text{r2}}{(\text{r2}-\text{r1})^2 (1-u)^2}
+\frac{1-4 \text{r1}}{ (\text{r2}-\text{r1})^2 u^2}
\right . \nonumber \\
&-&\left. \frac{2 \text{r2}}{\text{  }(\text{r2}- \text{r1})(\text{r2}-1) (1-u)}
+\frac{2 (1-2 \text{r1})}{\text{  }(1-\text{r1}) (\text{r2}-\text{r1}) u}
\right) G(\rho )
\nonumber \\
&+&\left(-\frac{\text{r1}}{(\rho-1 )^2}+\frac{2 \text{r1}}{(\text{r2}-\text{r1})^2 u^2}\right) \log (1-\text{r1})
+\left( \frac{ r_1 - r_2 - \text{r1} \text{r2}}{\text{r2} (\rho-1 )^2}
\right . \nonumber \\
&+& \left .\frac{4}{\rho-1}
-\frac{r_1 - r_2 - r_2(r_1 + r_2)}{\text{r2} (\text{r2}-\text{r1})^2 (1-u)^2}
+\frac{4}{(\text{r2}-\text{r1}) (1-u)}
\right) \log (1-\text{r2})
\nonumber \\
&-&\left(
\frac{3(1+\text{r1})}{ (\rho -1)^2}
+\frac {1 + 3 \text{r2} + 8 \text{r1} - 10 \text{r1} \text{r2} - 3 \text{r1}^2 + \text{r1}^2 \text{r2}}
{(1-\text{r1}) (\text{r2}-1) (\rho -1)}
+\frac{\text{r2}^2+\text{r1} \text{r2}+\text{r2}-\text{r1}}{\text{r2} (\text{r2}-\text{r1})^2 (1-u)^2}
\right . \nonumber \\
&+&\left . \frac{2 \text{r1}}{(\text{r2}-\text{r1})^2 u^2}
+\frac{5 \text{r2}^2+\text{r1} \text{r2}-2 \text{r2}+\text{r1}+1}{r_2 (\text{r2}-\text{r1}) (\text{r2}-1) (1-u)}
-\frac{1 - 3 r_1 - 4 \text{r1}^2}{ r_1(1-\text{r1}) (\text{r2}-\text{r1}) u }
\right . \nonumber \\
&+&\left . \frac{\text{r1}}{\rho ^3}
+\frac{2 \text{r1}-\text{r2} - 3 r_1 r_2}{2\text{  }\text{r2} \rho ^2}
+ \frac{\text{r2} \text{r1}^2-\text{r1}^2-\text{r2} \text{r1}-\text{r1}+\text{r2}}{\text{r1}\text{r2} \rho}
\right) \log (1-\rho )
\nonumber \\
&-&\left(\frac{6 (1+\text{r1})}{ (\rho-1)^3}
 +\frac{5 \text{r1}-3}{2(\rho-1)^2}
+\frac{1 + r_2 + 3 r_1 -4 \text{r1} \text{r2}-\text{r1}^2}{(1-\text{r1}) (\text{r2}-1) (\rho-1)}
\right . \nonumber \\
&+& \left . 
\frac{1 + \text{r1}+\text{r2}}{(\text{r2}-\text{r1}) (\text{r2}-1) (1-u)}
-\frac{1-4 \text{r1}}{(1-\text{r1}) (\text{r2}-\text{r1}) u}
\right) \log \left(\frac{m_b^2}{\mu ^2}\right)
\nonumber \\
&-&\frac{1 -  4 r_1 - r_2 + 3 r_1 r_2 + 2 r_1^2 - r_1^2 r_2 }{(1-\text{r1}) (\text{r2}-1) (\rho-1)}
+\frac{\text{r1}}{\text{r2}(\text{r2}-\text{r1}) (\text{r2}-1) (1-u)}
\nonumber \\
&-& \frac{1-2 \text{r1}}{(1-\text{r1}) (\text{r2}-\text{r1}) u}
+\frac{8 (1+\text{r1})}{(\rho-1 )^3}
-\frac{2 (1-2\text{r1})}{(\rho-1)^2}
-\frac{\text{r1}}{\rho ^2}
+\frac{2 \text{r1} (\text{r2}-1) +\text{r2}}{2\text{  }\text{r2} \rho }\, , 
\label{eq:t1sigma}
\ea

\ba
\frac{3}{\pi}{\rm Im}_s T_1^{\sigma} &=& \theta (1-\rho )
\left[ 
\left ( 1 + r_1 L_2(r_2) \right ) \left ( \frac{1}{\rho-1} \right )_+ \frac{d}{d\rho}  
- 2 \left ( 1 + \frac{ r_2 L_2(r_2)}{r_2-1} \right ) \left ( \frac{1}{\rho-1} \right )_+ 
\right . \nonumber \\
&& \left . \qquad - \left ( \frac{1 +r_1 + r_2}{(\rho-r_2)^2} - \frac{2 r_2}{(r_2-1)(\rho-r_2)} \right ) L_2(r_2) 
- \frac{1 + r_2}{r_2(\rho-r_2)^2} 
+ \frac{2}{\rho-r_2} 
\right ]
\nonumber \\
&& + \theta (\rho -1)\left[
\left ( (1+\text{r1}) \left ( \frac{  1- L_1(\rho) }{\rho-1}  \right )_+  
+ ( 1 + r_1 L_2(r_2)) \left ( \frac{1}{\rho-1} \right )_+ \right )\frac{d}{d\rho}
\right . \nonumber \\
&& \left . 
\qquad  +2 \left ( 3 + \frac{1}{r_2-1} - \frac{1}{1-r_1} \right )\left ( \frac{L_1(\rho)}{\rho-1} \right )_+
\right . \nonumber \\
&& \left . 
\qquad +\left ( -\frac{2 r_2}{r_2-1} L_2(r_2) + \frac{1+ r_1}{\rho} +\frac{2 (2 + r_1 )}{r_2-1} 
+ \frac{1 - 8 r_1 + r_1^2}{1-r_1} \right )\left ( \frac{1}{\rho-1} \right )_+
\right . \nonumber \\
&& \left . 
\qquad - \left ( \frac{1 +r_1 + r_2}{(\rho-r_2)^2} - \frac{2 r_2}{(r_2-1)(\rho-r_2)} \right ) L_2(r_2) 
\right . \nonumber \\
&& \left . 
\qquad - \left(\frac{1-4 \text{r1}}{(\text{r1}-\rho )^2}
+2 \frac{1-2 \text{r1}}{(\text{r1}-1) (\text{r1}-\rho)}
+\frac{2 \text{r2}}{(\text{r2}-1) (\rho -\text{r2})}-\frac{1+ \text{r1}+\text{r2}}{(\text{r2}-\rho )^2}\right) L_1(\rho) 
\right . \nonumber \\
&& \left .  
\qquad +\frac{\text{r1}}{\rho ^3} +\frac{2 r_1}{(r_1 -\rho)^2} 
-\frac{ 3 \text{r2}^2+\text{r1} \text{r2}+\text{r1}+1}{r_2 (\text{r2}-1) (\rho -\text{r2})}
+\frac{(\text{r2}-1) (1+ \text{r1}+\text{r2})}{\text{r2} (\text{r2}-\rho )^2}
\right . \nonumber \\
&& \left . 
\qquad -\frac{\text{r2}+\text{r1} (3 \text{r2}-2)}{2 \text{r2} \rho ^2}
- \frac{ \text{r1} (-\text{r2} \text{r1}+\text{r1}+\text{r2}+1)- \text{r2}}{\text{r1} r_2 \rho }
+\frac{ (1+\text{r1}) (1-4 \text{r1})}{r_1 (1-\text{r1}) (\text{r1} - \rho) }
\right]
\nonumber \\
&& +\delta (\rho -1) 
\left[
\frac{\pi ^2}{3} \left(\frac{1}{2}(1- 4 \text{r1}) \frac{d}{d\rho}
+  \frac{3 - 2 r_1}{1-\text{r1}}+\frac{4}{\text{r2}-1}  \right )
\right . \nonumber \\
&& \left . 
\quad +  \Big[ 
\log^2 (1-r_1) - (1 - 2 r_1) \log^2 (r_2-1)  + {\rm Li}_2(\text{r1}) -(1 + 2 r_1) {\rm Li}_2(1-\text{r2})
 \right . \nonumber \\
&&  \left . 
\quad - \left ( 2-r_1 - \frac{r_1}{r_2} + 2 r_1 \log (r_2-1) - r_1 \log r_2 \right) \log (1-r_1)
 \right . \nonumber \\
&&  \left . 
\quad +2 \left ( 2 + r_1 - \frac{r_1}{r_2} -  r_1 \log r_2 \right ) \log (r_2 -1) + 2(2 -r_1)
 \right . \nonumber \\
&&  \left .
\quad + \left ( - \frac{5 - 3 r_1}{2} + (1 - r_1) \log \frac{1-r_1}{r_2-1}  \right )\log \left(\frac{m_b^2}{\mu^2}\right) 
\Big]  \frac{d}{d\rho}
\right .  \nonumber \\
&& \left . 
\qquad -2 \left(2-\frac{1}{1-\text{r1}}\right) \log ^2(1-\text{r1})
-2 \left(\frac{1}{1- \text{r1}}+\frac{2}{ \text{r2}-1}\right) \log ^2(\text{r2}-1)
\right . \nonumber \\
&& \left . 
\qquad -2 \left(2-\frac{1}{1-\text{r1}}\right)  {\rm Li}_2(\text{r1}) 
+ 2 \left ( 3 - \frac{1}{1-r_1} + \frac{r_2+1}{r_2 -1} \right ) {\rm Li}_2(1-\text{r2}) 
\right . \nonumber \\
&& \left . 
\qquad - 2 + \text{r1} \left(1-\frac{1}{\text{r2}-1}\right) +\frac{1}{1-\text{r1}}
\right . \nonumber \\
&& \left . 
\qquad - 2 \left( -3  + \frac{1}{r_2-1} + \frac{1}{1-r_1} +  \frac{r_2}{r_2-1} \log r_2 -  \frac{2 r_2}{r_2-1}\log (r_2-1) \right )
\log (1-\text{r1})
\right . \nonumber \\
&& \left . 
\qquad 
- 2 \left( \frac{2}{1 - r_1} - \frac{3 + r_1}{r_2 -1} - \frac{2 r_2}{r_2-1} \log r_2 \right ) \log (r_2 -1)
\right . \nonumber \\
&& \left . 
\qquad +\left( 4 -\frac{3}{1-\text{r1}}+\frac{2+\text{r1}}{\text{r2}-1}
- 2 \left(2-\frac{\text{r2}}{\text{r2}-1}-\frac{1}{1-\text{r1}}\right) \log (1-\text{r1})
\right . \right . \nonumber \\
&& \left . \left .
\qquad  - 2 \left(\frac{1}{\text{r2}-1}+\frac{\text{r1}}{1-\text{r1}}\right) \log (\text{r2}-1)
\right) \log \left(\frac{m_b^2}{\mu^2}\right)
\right . \nonumber \\
&& \left . 
\qquad - (1+\text{r1}) \left(4-3 \log \left(\frac{m_b^2}{\mu ^2}\right)\right) 
\left ( \frac{d^2}{d\rho^2} - \delta (r_2-1) \frac{d}{d \rho} \right )
\right]\,.
\ea

As already mentioned, the above formulae are obtained 
in the $\overline{MS}$ scheme. To switch 
to the one-loop pole mass of $b$ quark the following 
expressions 
\ba
\Delta T_1 &=& - \frac{2\text{  }\rho  \left(3 \log \left(\frac{m_b^2}{\mu ^2}\right)-4\right)^{}}{(\rho -1)^2}\,,
\\
\Delta T_1^{p} &=& \frac{ (\text{r1}-\rho ) (\rho +1) \left(3 \log \left(\frac{m_b^2}{\mu ^2}\right)-4\right) }
{ (\text{r2}-\text{r1})(\rho -1)^2}\,,
\\
\Delta T_1^{\sigma} &=& \frac{ ((3-2 \rho ) \rho +\text{r1} (\rho +3)+3) \left(3 \log \left(\frac{m_b^2}{\mu ^2}\right)-4\right)\text{  }}
{6\text{  }(\rho -1)^3}\,.
\label{eq:TLOplus}
\ea
have to be added to the hard-scattering amplitudes $T_1$, $T_1^p$ and $T_1^\sigma$.
respectively. The corresponding additions to the imaginary parts are 
\ba
\frac{1}{2\pi}{\rm Im}_s \Delta T_1 &=& \delta(\rho-1)  \left(3 \log \left(\frac{m_b^2}{\mu ^2}\right)-4\right)
\left( 1 + \frac{d}{d\rho} \right)\,,
\\
\frac{r_2-r_1}{2 \pi}{\rm Im}_s \Delta T_1^{p} &=& \delta(\rho-1) 
\left(3 \log \left(\frac{m_b^2}{\mu ^2}\right)-4\right)\left( \frac{3 -r_1}{2} + (1-r_1)\left ( \frac{d}{d\rho} - \delta(r_2 -1)  \right )\right)\,, 
\nonumber \\
\\
\frac{3}{\pi}{\rm Im} \Delta_s T_1^{\sigma} &=& \delta(\rho-1)\left(3 \log \left(\frac{m_b^2}{\mu ^2}\right)-4\right) 
\left ( 1 +\frac{1-r_1}{2}\frac{d}{d\rho} 
\right . \nonumber \\
&& \left . \qquad\qquad  - (1+r_1)\left ( \frac{d^2}{d\rho^2} - \delta (r_2-1) \frac{d}{d \rho} \right )\right )\,.
\label{eq:IMTNLOplus}
\ea
 
\subsection{Amplitudes for $(f_{B\pi}^+  + f_{B\pi}^-)$ LCSR}

\ba
\widetilde{T}_1 &=&  
\frac{r_1^2 - r_1 r_2 -(1-\text{r1})(\text{r2}-\text{r1}) \log (1-\text{r1})}{r_1^2 (\rho -1)}
+\frac{(1-\text{r1}) (\text{r1}+\text{r2}) \log (1-\text{r1})}{\text{r1}^2 (\text{r2}-\text{r1}) u}
\nonumber \\
&&
-\frac{2 (\text{r2}-1) \log (1-\text{r2})}{\text{r2} (1-u) (\text{r2}-\text{r1})}
+\frac{(\rho -1) (\text{r2}+\rho ) \log (1-\rho )}{(\text{r2}-\text{r1}) (1-u) u \rho ^2}
+ \frac{\text{r2}-\text{r1}}{\text{r1} \rho }\,,
\ea


\ba
\frac{1}{\pi}{\rm Im}_s \widetilde{T}_1 &=&\theta (1-\rho )
\left[ \frac{2 (\text{r2}-1) }{\text{r2} (\text{r2}-\rho )} \right ]
\nonumber \\
&& + \theta (\rho-1) \frac{1}{r_1 - \rho} \left [
\frac{r_1 - r_2}{\rho^2} - \frac{(2 - r_2)(r_2 - r_1)}{r_2 \rho} + \frac{2 (r_2 -1)}{r_2}
\right ]
\nonumber \\
&& + \delta (\rho -1) \left[\frac{\text{r2}}{\text{r1}}+
\frac{(\text{r1}-1) (\text{r1}-\text{r2}) \log (1-\text{r1})}{\text{r1}^2} -1 \right ]\,,
\ea

\ba
\widetilde{T}_1^{p} 
&&=
\frac{4 G(\text{r1})}{(\text{r2}-\text{r1}) u}
+\frac{2 (\text{r2}-1) G(\text{r2})}{ (\text{r2}-\text{r1}) (1-u) (\rho -1)}
-\left(\frac{4}{ (\text{r2}-\text{r1}) u}+\frac{2}{ \rho-1 }+\frac{2}{ (\text{r2}-\text{r1}) (1-u)}\right) G(\rho )
\nonumber \\
&&
+\left(\frac{2 (\text{r1}+1)}{\text{r1} (\text{r2}-\text{r1}) u}-\frac{1- 2 r_1 -\text{r1}^2}{ \text{r1}^2 (\rho -1)}\right) 
\log (1-\text{r1})
+\left(\frac{2(\text{r2}-1)}{ \text{r2} (\rho-1)}+\frac{2 (\text{r2}-1)}{(\text{r2}-\text{r1}) \text{r2} (1-u)}\right) 
\log (1-\text{r2})
\nonumber \\
&& 
+\frac{2 (\rho +2) }{ (\rho -1)^2} \log\left (\frac{m_b^2}{\mu ^2}\right )
+\left(-\frac{2 (\text{r1}+1)}{\text{r1} (\text{r2}-\text{r1}) u}-\frac{2 (\text{r2}-1)}{(\text{r2}-\text{r1})
\text{r2} (1-u)}+\frac{2}{\rho -1}+\frac{\text{r2} \text{r1}+2 \text{r1}+2 \text{r2}}{\text{r1} \text{r2} \rho }
\right . \nonumber \\
&&
\left . \qquad -\frac{4}{\rho }+\frac{1}{\rho ^2}\right) \log (1-\rho )
+\frac{1}{ \rho }
- \frac{1+ 3 \text{r1}}{ \text{r1} (\rho-1 )}
-\frac{8}{ (\rho-1)^2}\,,
\ea

\ba
\frac{1}{\pi}{\rm Im}_s\widetilde{T}_1^{p} &=&\theta (1-\rho )
\left[ 2 \frac{r_2 -1}{\rho-r_2} L_2(r_2) \left (\frac{1}{\rho-1} \right )_+ \right ]
+ \theta (\rho-1)  \left [ 
- 2 \left(1 +\frac{ 1 -\text{r2}}{\rho -\text{r2}} L_2(r_2)\right)\left(\frac{1}{ \rho -1}\right)_+
\right .\nonumber \\ 
&& \left . 
+2\left(\frac{ L_1(\rho)}{\rho -1}\right)_+  
+ \frac{2 (-\text{r1}+2 \text{r2}-\rho )}{(\text{r1}-\rho ) (\rho -\text{r2})} L_1(\rho)
-\frac{2 \text{r1}+2 \text{r2}-3 \text{r2} \text{r1}}{\text{r1} \text{r2} \rho } 
+ 2 \frac{(r_1+1)}{r_1(\rho-r_1)} 
\right . \nonumber \\
&& \left . 
- 2 \frac{r_2-1}{r_2 (\rho-r_2)}
-\frac{1}{\rho ^2} \right]
\nonumber \\
&& + \delta(\rho-1) 
\left[ 
2 \left(\log (\text{r2})+\frac{2}{\text{r2}}-3\right) \log (\text{r2}-1)
 + \left (\frac{1}{r_1^2} - \frac{\text{r1} (2-\text{r1})}{\text{r1}^2} -\frac{2}{\text{r2}}\right) \log (1 -\text{r1}) 
\right . \nonumber \\
&& \left . - 2 \log \left (\frac{ \text{r2}-1}{1-\text{r1}} \right ) L_1(r_2) + 4 {\rm Li}_2(1-r_2)
+\frac{1}{\text{r1}}+3 
- 2 \log \left ( \frac{m_b^2}{\mu^2} \right ) + \frac{4}{3} \pi^2
\right . \nonumber \\
&& \left . + 2 \left ( 4 - 3 \log \left ( \frac{m_b^2}{\mu^2} \right ) \right ) 
\left ( \frac{d}{d\rho} - \delta (r_2 -1) \right ) 
\right ]\,,
\nonumber \\
\ea

\ba
\widetilde{T}_1^{\sigma} &=&  
\frac{2 (\text{r1}-1) G(\text{r1})}{3\text{ }u^2 (\rho -1) (\text{r1}-\text{r2})}
+\left(\frac{\text{r2}-\text{r1}}{3\text{  }(\rho-1)^2}-\frac{1}{3\text{  }(\text{r2}-\text{r1}) (1-u)^2}\right) G(\text{r2})
\nonumber \\
&&+\left(\frac{2 (\text{r2}-\text{r1})}{3\text{  }(\text{r1}-1) (\rho-1)}
+\frac{2}{3\text{  }(1-\text{r1}) u}
-\frac{\text{r2}-\text{r1}}{3\text{  }(\rho-1)^2}
+\frac{1}{3 (1-u)^2 (\text{r2}-\text{r1})}
\right . \nonumber \\
&& \left . \qquad +\frac{2}{3\text{ }u^2 (\text{r2}-\text{r1})}\right) G(\rho )
\nonumber \\
&& +\left(\frac{1+\text{r1}}{3\text{  }\text{r1} (\text{r1}-\text{r2}) u^2}+\frac{1}{3\text{  }\text{r1}^2 u}
+\frac{\text{r2}-\text{r1}}{3\text{  }\text{r1}^2 (1-\rho )}
-\frac{\text{r1}^3-\text{r2} \text{r1}^2+\text{r1}-\text{r2}}{6 \text{r1}^2 (\rho-1)^2}\right) \log (1-\text{r1})
\nonumber \\
&& +\left(\frac{\text{r2}^2-\text{r1} \text{r2}-\text{r2}+\text{r1}}{3\text{r2} (\rho-1)^2}
-\frac{\text{r2}-1}{3\text{ }\text{r2} (\text{r2}-\text{r1}) (1-u)^2}\right) \log (1-\text{r2})
\nonumber \\
&&+\left(\frac{ 1-\text{r1}}{3 \text{r1} (\text{r2}-\text{r1}) u^2}
+\frac{ 1+\text{r2}}{3(\text{r2}-1) \text{r2} (1-u)}
+\frac{ (\text{r2}-\text{r1}) \left((\text{r2}-1) \text{r1}^2-2 \text{r2} \text{r1}+\text{r2}\right)}{3\text{r1}^2 \text{r2} \rho }
\right . \nonumber \\
&& \left . +\frac{1}{3(\text{r2}-\text{r1}) (1-u)^2}
-\frac{1}{3\text{r2} (\text{r2}-\text{r1}) (1-u)^2}
+\frac{2}{3(\text{r2}-\text{r1}) u^2}
+\frac{ \text{r2}-\text{r1}}{(\rho -1)^2}
\right . \nonumber \\
&& \left . +\frac{(\text{r2}-\text{r1}) (2- 3 \text{r2})}{6\text{r2} \rho^2}
+\frac{ \text{r2}-\text{r1}}{3\rho^3}
+\frac{3 \text{r1}-1}{3\text{r1}^2 u (1-\text{r1})}
+\frac{2}{3u (1-\text{r1})}
+\frac{2 (\text{r2}-\text{r1})}{3(\rho -1) (\text{r1}-1)}
\right . 
\nonumber \\
&& \left . +\frac{ (\text{r1} (\text{r2}-3)-3 \text{r2}+5) (\text{r2}-\text{r1})}{3(\text{r2}-1) (\rho -1) (1-\text{r1})}\right) 
\log (1-\rho)
\nonumber \\
&&+\left(\frac{2 (\text{r2}-\text{r1})}{ (\rho-1)^3}+\frac{1}{3\text{  }(\text{r2}-1)(1-u)}
+\frac{2}{3\text{  }(1-\text{r1}) u}
\right . \nonumber \\
&& \left . \qquad +\frac{-\text{r1}^2-\text{r2} \text{r1}+3 \text{r1}+2 \text{r2}^2-3 \text{r2}}{3\text{  }(\text{r1}-1)
(\text{r2}-1) (\rho-1 )}
+\frac{2 (\text{r2}-\text{r1})}{3\text{  }(\rho-1)^2}\right) \log \left(\frac{m_b^2}{\mu ^2}\right)
\nonumber \\
&& +\frac{1}{3 r_2 (1-\text{r2}) (1-u)}
+\frac{1}{3\text{  }\text{r1} u}
+\frac{-\text{r2} \text{r1}^2+2 \text{r1}^2+\text{r2}^2 \text{r1}-\text{r2} \text{r1}-\text{r1}-\text{r2}^2+\text{r2}}
{3\text{  }\text{r1} (\text{r2}-1) (\rho-1)}
\nonumber \\
&& +\frac{-\text{r2}^2+\text{r1} \text{r2}+\text{r2}-\text{r1}}{3\text{ }\text{r2} \rho }
-\frac{-7 \text{r1}^2+7 \text{r2} \text{r1}+\text{r1}-\text{r2}}{6\text{  }\text{r1} (\rho-1 )^2}
+ \frac{\text{r2}-\text{r1}}{3\rho ^2}+\frac{8 (\text{r2}-\text{r1})}{3\text{  }(1-\rho )^3}\,,
\ea

\ba
\frac{3}{r_2-r_1}\frac{1}{\pi}{\rm Im}_s\widetilde{T}_1^{\sigma} &=&\theta (1-\rho )
\left [  \frac{L_2(r_2)}{(r_2 - \rho)^2} - L_2(r_2)\left ( \frac{1}{\rho-1} \right )_+  \frac{d}{d\rho}
\right ]
\nonumber \\
&& +\theta (\rho-1) 
\left [  \left (\frac{L_1(\rho)- L_2(r_2)-1}{\rho-1} \right )_+   \frac{d}{d\rho} 
+\frac{L_2(r_2)}{(\text{r2}-\rho )^2}
\right . \nonumber \\
&& \left . 
+ \left(-\frac{2}{(\text{r1}-\rho )^2}-\frac{1}{(\text{r2}-\rho )^2}
+\frac{2}{(\text{r1}-1) (\rho -\text{r1})}\right) L_1(\rho)
\right . \nonumber \\
&& \left . 
- 2 \left ( \frac{3 - r_1 - 2 r_2}{(1-r_1)(r_2-1)}  \right ) \left ( \frac{1}{\rho-1} \right )_+
+  \frac{2} {1 -r_1} \left ( \frac{L_1(\rho)}{\rho-1} \right )_+ 
\right . \nonumber \\
&& \left . 
+ \frac{1+r_2}{r_2 (r_2-1)(\rho-r_2)} + \frac{r_1^2 + 2 r_1 r_2 - r_2}{r_1^2 r_2 \rho}
\right . \nonumber \\
&& \left . 
+ \frac{ r_1 ( 2 r_1+3) -1}{(r_1-1) r_1^2 (\rho-r_1)}
+ \frac{3 \text{r2}-2}{2 \text{r2} \rho ^2}
- \frac{1 +r_1}{ \text{r1} (\rho -\text{r1})^2 }
+\frac{1-r_2}{\text{r2} (\rho -\text{r2})^2 }-\frac{1}{\rho^3}
\right ]
\nonumber \\
&& + \delta (\rho -1) \left[
\frac{\pi ^2}{3} \left(2 \frac{d}{d\rho}-\frac{1}{1-\text{r1}}\right)
- \Bigg (  2 \log (r_2-1) \left ( 1 - \log (r_2 -1) 
 \right.
\right . \nonumber \\
&&  \left.\left .
+ L_2(r_2) \right ) 
+ \log \left(\frac{m_b^2}{\mu ^2} \right )\left ( 1 - \log (r_2-1) \right ) - 2 {\rm Li}_2(1 - r_2)
 \right . \nonumber \\
&& \left .  
+ \frac{1}{2} \log (1-r_1) \left ( 1 + \frac{1}{r_1^2} - 2 L_2(r_2) \right ) 
+ \frac{1}{2} \left (\frac{1}{r_1} -3 \right ) \Bigg )\frac{d}{d\rho} 
\right. \nonumber \\ 
&& \left . +\frac{2 \log (1-\text{r1})}{1-\text{r1}} \left ( 1 + \frac{1-r_1}{2 r_1^2} - \log (1 -r_1) \right )
\right. \nonumber \\ 
&& \left . +\frac{2 \log (\text{r2}-1)}{1- \text{r1}} \left ( \log (r_2-1) + \frac{r_1 + r_2 - 2}{r_2 -1} \right )
+  \frac{1}{r_2-1} + \frac{1}{r_1} -1 
\right . \nonumber \\ 
&& \left . +\frac{2}{\text{r1}-1}\left(
-\frac{ r_1 + 2 r_2 -3}{ 2 (\text{r2}-1)} + \log (1-\text{r1}) - \log (\text{r2}-1) \right) 
\log \left(\frac{m_b^2}{\mu ^2}\right)
\right . \nonumber \\
&& \left . +\frac{2}{\text{r1}-1} \left ({\rm Li}_2(\text{r1}) - {\rm Li}_2(1-\text{r2}) \right )
\right . \nonumber \\
&& \left . 
+ \left(4-3 \log \left(\frac{m_b^2}{\mu ^2}\right)\right) 
\left ( \frac{d^2}{d\rho^2} - \delta (r_2 -1) \frac{d}{d \rho} \right )
\right ]\,,
\ea


\ba
\Delta \widetilde{T}_1&=& 0\,,
\\
\Delta \widetilde{T}_1^{p} &=& -\frac{ (\rho +1) \left(3 \log \left(\frac{m_b^2}{\mu ^2}\right)-4\right) }{ (\rho -1)^2}\,,
\\
\Delta \widetilde{T}_1^{\sigma} &=& -\frac{(\text{r2}-\text{r1})\text{  }(\rho +3) \left(3 \log \left(\frac{m_b^2}{\mu ^2}\right)-4\right)}
{6\text{  }(\rho -1)^3}\,,
\label{eq:TNLOplusminus}
\ea


\ba
\frac{1}{\pi}{\rm Im}_s \Delta \widetilde{T}_1^{p} &=&  \delta(\rho -1) \left(3 \log \left(\frac{m_b^2}{\mu ^2}\right)-4\right) 
\left(1 + 2 \left (\frac{d}{d\rho} - \delta (r_2-1) \right ) \right)\,,
\\
\frac{3}{r_2-r_1}\frac{1}{\pi}{\rm Im}_s\widetilde{T}_1^{\sigma} &=&  \delta(\rho-1) 
\left(3 \log \left(\frac{m_b^2}{\mu ^2}\right)-4\right)\left(\frac{1}{2}\frac{d}{d\rho} 
+ \frac{d^2}{d\rho^2} - \delta(r_2-1) \frac{d}{d\rho} \right ).
\label{eq:IMTNLOplusminus}
\ea
\vspace*{1cm}
\subsection{Amplitudes for $f_{B \pi}^T$ LCSR}
\ba
\frac{1}{2} T_1^{T} &=&  
\left(\frac{ 1-\text{r1}}{(\text{r2}-\text{r1}) (\rho-1 )}-\frac{ \text{r2}-1}{(\text{r2}-\text{r1})^2 u}
+\frac{ \text{r2}-1}{(\text{r2}-\text{r1}) (\rho-1 )}\right) G(\text{r1})
\nonumber \\
&&+ \left(\frac{ 1-\text{r1}}{(\text{r2}-\text{r1})^2 (1-u)}+\frac{1}{\rho-1}\right) G(\text{r2})
+\left(-\frac{1-\text{r1}}{(\text{r2}-\text{r1})^2 (1-u)}
+\frac{\text{r2}-1 }{(\text{r2}-\text{r1})^2 u}-\frac{2 }{\rho-1}\right) G(\rho )
\nonumber \\
&&+\left(-\frac{1- \text{r1}}{\text{r1} (\text{r2}-\text{r1}) u}+\frac{ 1-\text{r1}}{\text{r1} (\rho-1 )}\right) \log (1-\text{r1})
\nonumber \\
&&+\left(\frac{\text{r2}-1}{(\text{r2}-\text{r1}) \text{r2} (1-u)}+\frac{ \text{r2}-1}{\text{r2} (\rho-1 )}\right) 
\log (1-\text{r2})
\nonumber \\
&&+ \left(-\frac{\text{r2}-1 }{(\text{r2}-\text{r1}) \text{r2} (1-u)}+\frac{1}{2\rho ^2}
+\frac{1-\text{r1}}{\text{r1} (\text{r2}-\text{r1}) u}-\frac{-2 \text{r1}+ \text{r2} \text{r1}+2\text{  }\text{r2}}{2 \text{r1} \text{r2}
\rho }\right) \log (1-\rho )
\nonumber \\
&&+ \left(- \frac{1}{2 (\rho-1)} +\frac{3 }{(\rho-1 )^2}\right) \log \left(\frac{m_b^2}{\mu ^2}\right)
+\frac{1}{\rho-1 }+\frac{1}{2 \rho } -\frac{4 }{(\rho-1 )^2}\,,
\ea

\ba
\frac{1}{2 \pi}{\rm Im}_s T_1^{T} &=&\theta (1-\rho )
\left [  
\frac{1-r_1}{(r_2-r_1)(\rho -r_2)}L_1(r_2) -  L_2(r_2)\left (\frac{1}{\rho-1} \right )_+  + \frac{r_2-1}{r_2 (\rho - r_2) }
\right ]
\nonumber \\
&& + \theta(\rho-1) \left [ 
\frac{1-r_1}{(r_2-r_1)(\rho -r_2)}L_1(r_2) + \frac{r_1+r_2 -\rho -1}{(r_1-\rho)(r_2 -\rho)}L_1(\rho) 
\right . \nonumber \\
&& \left . -  L_2(r_2) \left (\frac{1}{\rho-1} \right )_+  + 2 \left ( \frac{L_1(\rho)}{\rho-1} \right )_+ 
+ \frac{r_1-1}{r_1(\rho-r_1)} + \frac{2(r_2-r_1) + r_1 r_2}{2 r_1 r_2 \rho} - \frac{1}{2 \rho^2}
\right ]
\nonumber \\
&& + \delta (\rho -1) \left[
-\left ( \log  \frac{\text{r2}-1}{1 - r_1} \right )^2
+\left(-\log (\text{r2})-\frac{1}{\text{r1}}-\frac{1}{\text{r2}}+2\right) \log (1-\text{r1})
\right . \nonumber \\
&& \left . + 2\log (\text{r2}-1) \left(\log (\text{r2})+\frac{1}{\text{r2}}-1\right)
+\frac{1}{2} \log \left(\frac{m_b^2}{\mu ^2}\right)-{\rm Li}_2(\text{r1})+ 3 \, {\rm Li}_2(1-\text{r2})
\right . \nonumber \\
&& \left . -1 +\frac{\pi^2}{2}
+ \left( 4 - 3 \log \left(\frac{m_b^2}{\mu ^2}\right)\right) \frac{d}{d\rho} 
\right]\,,
\ea

\ba
\frac{1}{2}T_1^{T\,p} &=&  \left(- \frac{3}{(\text{r2}-\text{r1})^2 u} +\frac{1}{(\text{r2}-\text{r1}) (\rho-1 )}\right) G(\text{r1})
\nonumber \\
&& -\left(\frac{1}{(\text{r2}-\text{r1}) (\rho-1)}+\frac{3}{(\text{r2}-\text{r1})^2 (1-u)}\right) G(\text{r2})
+\frac{3 G(\rho )}{(\text{r1}-\text{r2})^2 u (1-u)}
\nonumber \\
&& +\left(\frac{1-2 \text{r1}}{\text{r1} (\text{r2}-\text{r1}) (\rho -1)}-\frac{2}{(\text{r2}-\text{r1})^2 u}\right) \log (1-\text{r1})
\nonumber \\
&&+\left(\frac{1}{(\text{r2}-\text{r1}) \text{r2} (\rho -1)}-\frac{2}{(\text{r2}-\text{r1})^2 (1-u)}\right) \log (1-\text{r2})
\nonumber \\
&& +\left(\frac{2}{(\text{r2}-\text{r1}) \rho }+\frac{2}{(\text{r2}-\text{r1})^2 u (1-u) }\right) \log (1-\rho )\,,
\ea

\ba
\frac{r_2-r_1}{2 \pi}{\rm Im}_s T_1^{T\,p} &=&\theta (1-\rho )
\left [ \frac{3}{r_2 -\rho} L_1(r_2) + (L_2 (r_2) - 1) \left (\frac{1}{\rho-1} \right )_+  + \frac{2}{r_2 -\rho}
\right] 
\nonumber \\
+ && \!\!\!\!\!\!\theta(\rho-1) 
\left [ \frac{3}{r_2 -\rho} L_1(r_2) + \frac{3(r_1-r_2)}{(r_2-\rho)(\rho-r_1)} L_1(\rho) 
\right . \nonumber \\
&& \left .
+ (L_2 (r_2) - 1) \left (\frac{1}{\rho-1} \right )_+  - \frac{2 (r_1- 2\rho)}{(r_1-\rho)\rho }
\right ]
\nonumber \\
- && \delta (\rho -1) 
\left[ \log^2(1-\text{r1})+\left(2 \log (\text{r2}-1)-\log (\text{r2})+\frac{1}{\text{r1}}-\frac{1}{\text{r2}}-4\right) \log (1-\text{r1})
\right .  \nonumber \\
&& \left . -3 \log ^2(\text{r2}-1)+ 2 \log (\text{r2}-1) \left(\log (\text{r2})+\frac{1}{\text{r2}}+1\right)
\right .  \nonumber \\
&& \left . -2 \left ( \log \frac{ \text{r2}-1}{1 - r_1} \right ) \log \left(\frac{m_b^2}{\mu ^2}\right)
+{\rm Li}_2(\text{r1})+{\rm Li}_2(1-\text{r2}) +\frac{5 \pi ^2}{6}
\right]\,,
\ea

\ba
\frac{1}{2} T_1^{T\,\sigma} &=&  
\left(\frac{-1}{3 (1-\text{r1}) (\rho-1 )}-\frac{1}{6 (\rho-1)^2}
+\frac{1}{3 (1-\text{r1}) (\text{r2}-\text{r1}) u}+\frac{1}{2 (\text{r2}-\text{r1})^2 u^2}\right) G(\text{r1})
\nonumber \\
&& +\left(\frac{1}{3 (\text{r2}-1) (\rho-1 )}-\frac{1}{6 (\rho-1 )^2}
+\frac{1}{3 (\text{r2}-\text{r1}) (\text{r2}-1) (1-u)}
\right . \nonumber \\
&& \left . 
+\frac{1}{2 (\text{r2}-\text{r1})^2 (1-u)^2}\right) G(\text{r2})
\nonumber \\
&&+\left(\frac{\text{r1}+\text{r2}-2}{3 (1-\text{r1}) (\text{r2}-1) (\rho-1)}
-\frac{1}{3 (\text{r2}-\text{r1}) (\text{r2}-1) (1-u)}-\frac{1}{3 (1-\text{r1}) (\text{r2}-\text{r1}) u}
\right . \nonumber \\
&& \left . -\frac{1}{2 (\text{r2}-\text{r1})^2 (1-u)^2}-\frac{1}{2 (\text{r2}-\text{r1})^2 u^2}+\frac{1}{3 (\rho-1)^2}\right) G(\rho )
\nonumber \\
&&+\left(\frac{1}{3 (\text{r2}-\text{r1})^2 u^2}-\frac{1}{3 (\text{r2}-\text{r1}) u \text{r1}}+\frac{1}{3 (\rho-1) \text{r1}}
-\frac{1}{6 (\rho-1)^2 \text{r1}}\right) \log (1-\text{r1})
\nonumber \\
&& +\left(\frac{1- 2 \text{r2}}{6 \text{r2} (\rho-1 )^2}+\frac{1}{3 (\text{r2}-\text{r1}) \text{r2} (1-u)}
+\frac{1}{3 \text{r2} (\rho-1)}
+\frac{1}{3 (\text{r2}-\text{r1})^2 (1-u)^2}\right) \log (1-\text{r2})
\nonumber \\
&&+\left(\frac{\text{r1}+\text{r2}-2}{2 (1-\text{r1}) (\text{r2}-1) (\rho-1 )}-\frac{1}{2 (\text{r2}-\text{r1})
(\text{r2}-1) (1-u)}
-\frac{1}{2 (1-\text{r1}) (\text{r2}-\text{r1}) u}
\right . \nonumber \\
&& \left . -\frac{1}{(\rho-1 )^2}-\frac{2}{(\rho-1)^3}\right) \log \left(\frac{m_b^2}{\mu ^2}\right)
\nonumber \\
&& -\left(\frac{5 \text{r1}+1}{6 (1-\text{r1}) \text{r1} (\text{r2}-\text{r1}) u}
+ \frac{ 5 r_2 + 1}{ 6 (\text{r2}-1) \text{r2} (\text{r2}-\text{r1})(1-u)}
\right . \nonumber \\
&& \left . +\frac{\text{r2} \text{r1}-4 \text{r1}-4 \text{r2}+7}{3 (1-\text{r1}) (\text{r2}-1) (\rho-1 )}
-\frac{-2 \text{r2} \text{r1}+\text{r1}+\text{r2}}{6 \text{r1} \text{r2} \rho }
\right . \nonumber \\
&& \left . +\frac{1}{3 (\text{r2}-\text{r1})^2 (1-u)^2}+\frac{1}{3 (\text{r2}-\text{r1})^2 u^2}
+\frac{1}{(\rho-1 )^2}\right) \log (1-\rho )
\nonumber \\
& & + \frac{2 - \text{r1}-\text{r2}}{6 (1-\text{r1}) (\text{r2}-1) (\rho-1 )}
+ \frac{1}{6 (r_2-r_1) (r_2 -1) (1-u) }
+\frac{1}{6 (1-\text{r1}) (\text{r2}-\text{r1}) u}
\nonumber \\
&& +\frac{5}{3 (\rho-1 )^2}+\frac{8}{3 (\rho-1)^3}\,,
\ea


\ba
\frac{3}{2 \pi}{\rm Im}_s T_1^{T\, \sigma} &=&\theta (1-\rho )
\left[ \frac{1}{\rho-r_2} \left [ \left ( 
- \frac{3}{2 (\rho-r_2)} + \frac{1}{r_2-1} \right ) L_2(r_2) + \frac{r_2-3}{2 r_2 (\rho-r_2)} \right ]
\right . \nonumber \\
&& \left . - \left ( \frac{L_2(r_2)}{r_2-1} - \frac{1 + L_2(r_2)}{2} \frac{d}{d \rho} \right ) \left ( \frac{1}{\rho-1} \right )_+ \right ]
\nonumber \\
&& + \theta (\rho-1 ) 
\left [ \left ( \left (\frac{1- L_1(\rho)}{\rho-1} \right )_+ + \frac{1 + L_2 (r_2) }{2} \left ( \frac{1}{\rho-1} \right )_+ 
\right )  \frac{d}{d\rho} 
+ \frac{r_1 + r_2 -2}{ (r_1-1)(r_2-1)} \left ( \frac{L_1(\rho)}{\rho-1} \right )_+ 
\right . \nonumber \\
&& \left . 
+ \left ( - \frac{L_2 (r_2)}{ r_2-1} + \frac{4 -r_1 }{ r_1-1} + \frac{3}{r_2-1} + \frac{1}{\rho} \right ) 
\left ( \frac{1}{\rho-1} \right )_+ 
\right . \nonumber \\
&& \left . 
+ \left ( \frac{1}{\rho-r_2}\left ( \frac{3}{2 (\rho-r_2)} - \frac{1}{r_2-1} \right ) 
+ \frac{1}{\rho-r_1} \left ( \frac{3}{2 (\rho-r_1)} - \frac{1}{r_1-1} \right ) 
\right ) L_1(\rho)
\right . \nonumber \\
&& \left . + \left ( - \frac{3}{2 (\rho - r_2)} + \frac{1}{r_2-1} \right ) \frac{L_2(r_2)}{\rho - r_2}
\right . \nonumber \\
&& \left . - \frac{1}{2} \left ( \frac{1 + 5 r_1}{r_1(r_1-1)(\rho-r_1)} + \frac{1 + 5 r_2}{r_2(r_2-1)(\rho-r_2)} - \frac{2}{(\rho-r_1)^2} 
\right . \right . \nonumber \\
&& \left . \left .
- \frac{3(r_2-1)}{r_2 (\rho-r_2)^2} + \frac{1}{\rho} \left (\frac{1}{r_1} + \frac{1}{r_2} - 2 \right ) \right ) \right ] 
\nonumber \\
&& + \delta(\rho-1) \left [
\frac{\pi^2}{4} \left ( - \frac{d}{d\rho} + \frac{2}{3} \left ( \frac{1}{1-r_1} + \frac{4}{r_2-1} \right ) \right ) 
\right .  \nonumber \\
&& +  \left . 
\Big [ \frac{1}{2} \log^2 (1 - r_1)
+ \left ( - \log (r_2-1) + \frac{1}{2} \log (r_2) + \frac{ r_1 + r_2 - 2 r_1 r_2}{2 r_1 r_2 } \right ) \log (1 - r_1)
\right .  \nonumber \\
&&  \left . + \frac{1}{2} \log^2 (r_2 - 1) - \left (  \log r_2 +  \frac{1 - 3 r_2}{r_2} \right ) \log (r_2-1)
 \right . \nonumber \\
&&  \left . - 3 + 2 \log \left(\frac{m_b^2}{\mu^2}\right) 
+ \frac{1}{2} \left ( {\rm Li}_2(\text{r1}) - 3 \,  {\rm Li}_2(1-\text{r2}) \right )
\Big ] \frac{d}{d\rho}
\right . \nonumber \\
&& \left . 
+\frac{\log ^2(1-\text{r1})}{ 1-\text{r1}}
- \left (  -2 \frac{ \log (r_2 -1)}{r_2 -1} + \frac{\log r_2}{r_2-1} + \frac{1}{r_1 (1 - r_1)} + \frac{1}{r_2 (r_2-1)} \right) \log (1-\text{r1})
\right . \nonumber \\
&& \left . -
\left ( \frac{2}{r_2-1} + \frac{1}{1- r_1} \right ) \log ^2(\text{r2}-1) + \frac{r_1 + r_2 -2}{2 (1 -r_1) (r_2 -1)}
\right . \nonumber \\
&& \left . - 2 \left ( - \frac{2}{r_2 -1} + \frac{1}{r_2} + \frac{1}{1 -r_1} - \frac{\log r_2}{r_2-1}   \right ) \log \ (\text{r2}-1) 
\right . \nonumber \\
&& \left .
 -\left( \frac{ 3(r_1 + r_2 - 2)}{ 2 (1-r_1) (r_2-1)} 
+ \left ( \frac{1}{r_2 -1} + \frac{1}{1 - r_1} \right )\log \left (\frac{r_2-1}{1-r_1} \right ) \right )
\log \left(\frac{m_b^2}{\mu^2}\right)
\right . \nonumber \\
&& \left . +\frac{1}{1-r_1} \left ( {\rm Li}_2(\text{r1}) - \frac{2 r_1 + r_2 -3}{r_2 -1} {\rm Li}_2(1-\text{r2}) \right )
\right . \nonumber \\
&& \left . 
- \left(4-3 \log \left(\frac{m_b^2}{\mu^2}\right)\right) 
\left ( \frac{d^2}{d\rho^2} - \delta(r_2 -1) \frac{d}{d\rho} \right )
 \right ]\,,
\ea

\ba
\Delta T_1^{T} &=& -\frac{ (\rho +1) \left(3 \log \left(\frac{m_b^2}{\mu ^2}\right)-4\right) }{(\rho -1)^2}\,,
\\
\Delta T_1^{T\,p} &=& 0\,,
\\
\Delta T_1^{T \, \sigma} &=& \frac{ 2 (\rho +1) \left(3 \log \left(\frac{m_b^2}{\mu ^2}\right)-4\right) }{3 (\rho -1)^3}\,.
\label{eq:TNLOT}
\ea


\ba
\frac{1}{2 \pi}{\rm Im}_s \Delta T_1^{T} &=& \delta(\rho -1) \left(3 \log \left(\frac{m_b^2}{\mu ^2}\right)-4\right) 
\left( \frac{1}{2} + \frac{d}{d\rho} \right)\,,
\\
\frac{3}{2 \pi}{\rm Im}_s \Delta T_1^{T\,\sigma} &=& - \delta(\rho -1) \left(3 \log \left(\frac{m_b^2}{\mu ^2}\right)-4\right) 
\left( \frac{d}{d\rho} + \left ( \frac{d^2}{d\rho^2} - \delta(r_2-1) \frac{d}{d\rho} \right ) \right).
\label{eq:IMTNLOT}
\ea

\section{Two-point sum rule for $f_B$ }

We use the sum rule with $O(\alpha_s)$ accuracy 
with the perturbative part calculated in the $\overline{MS}$ scheme
for $b$-quark \cite{JL}:  
\ba
f_B^2 &=& \frac{e^{m_B^2/\overline{M}^2}}{m_B^4}\Bigg[
\frac{3m_b^2}{8\pi^2}\int\limits_{m_b^2} ^{\overline{s}_0^B}ds
e^{-s/\overline{M}^2}\Bigg \{
\frac{(s-m_b^2)^2}{s}+ \frac{\alpha_sC_F}{\pi}\rho_1(s,m_b^2)\Bigg \}
\nonumber
\\
&& +m_b^2e^{-m_b^2/\overline{M}^2}\Bigg \{-m_b \langle \bar q q \rangle \left(1+
\frac{\alpha_sC_F}{\pi}\delta_1(\overline{M}^2,m_b^2)+
\frac{m_0^2}{2\overline{M}^2}\left(1-\frac{m_b^2}{2\overline{M}^2}\right)\right)
\nonumber 
\\
&& + \frac1{12}\langle  \frac{\alpha_s}{\pi}
G G \rangle
 -\frac{16\pi}{27}\frac{\alpha_s\langle \bar q q \rangle^2}{\overline{M}^2}
\left(1-\frac{m_b^2}{4\overline{M}^2}-\frac{m_b^4}{12\overline{M}^4}\right)
\Bigg\}\Bigg]~,
\label{eq:fBSRMSbar}
\ea
where $\overline{M}$ and $\overline{s}_0^B$ are, respectively, 
the Borel parameter and
effective threshold. In the above, the functions determining the spectral density of the 
$O(\alpha_s)$ corrections to the perturbative and quark condensate
terms are
\ba
\rho_1(s,m_b^2)=\frac{s}2(1-x)\Bigg \{(1-x)\big[4 {\rm Li}_2(x)+
2\ln x \ln (1-x)-(5-2x)\ln (1-x)\big]
\nonumber\\
+(1-2x)(3-x)\ln x +3(1-3x)\ln \left ( \frac{\mu^2}{m_b^2} \right )+\frac12(17-33x)\Bigg \}\,,
\ea
where $x=\frac{m_b^2}{s}$, and 
\ba
\delta_1(\overline{M}^2,m_b^2)= -\frac32\Bigg[\Gamma \left (0,\frac{m_b^2}{\overline{M}^2} \right )
e^{m_b^2/\overline{M}^2}- 1-\Bigg (1-\frac{m_b^2}{\overline{M}^2} \Bigg ) \Bigg(\ln
\left (\frac{\mu^2}{m_b^2}\right )+\frac43\Bigg)\Bigg] , 
\label{eq:fBSRcorre}
\ea
respectively, 
and $\Gamma(n,z)$ is the incomplete $\Gamma$ function.

\end{document}